\documentclass[11pt,authoryear]{elsarticle}
\usepackage{amssymb}
\usepackage{amsmath}
\usepackage{amsmath,amsfonts,amssymb}
\usepackage{stmaryrd}
\usepackage{natbib}
\usepackage{graphicx}
\usepackage{color,xcolor}
\usepackage{appendix}
\usepackage{soul}
\usepackage[utf8]{inputenc}
\setcitestyle{authoryear,round}
\usepackage{marginnote}
\usepackage{hyperref}
\hypersetup{colorlinks=true, allcolors = {red}}
\usepackage{changes}
\definechangesauthor[name={XY}, color=magenta]{XY}
\makeatletter \oddsidemargin 0.0cm \evensidemargin 0.0cm
\def \Curl{\mbox{Curl\hskip 1pt}}

\def \Div{\mbox{Div\hskip 1pt}}

\def \Grad{\mbox{Grad\hskip 1pt}}
\def\rr#1{(\ref{#1})}
\def\bm#1{\mbox{\boldmath{$#1$}}}
\newcommand{\be}{\begin{equation}}
\newcommand{\en}{\end{equation}}
\newcommand{\la}{\label}
\newcommand{\ep}{\varepsilon}
\newcommand{\vep}{\epsilon}
\newcommand{\paa}{\partial}

\newcommand{\s}[1]{{\Large\textsf{\textbf{#1}}}}
\begin{document}
\begin{frontmatter}
\title{\s{Axisymmetric necking of a circular electrodes-coated dielectric membrane}}
\author[mymainaddress]{Yibin Fu\corref{mycorrespondingauthor}} \ead{y.fu@keele.ac.uk}
\cortext[mycorrespondingauthor]{Corresponding author}% at: School of Computing and Mathematics, Keele University, Staffs ST5 5BG, UK}
\author[secondaddress]{Xiang Yu}
\address[mymainaddress]{School of Computer Science and Mathematics, Keele University, Staffs ST5 5BG, UK}
\address[secondaddress]{School of Computer Science and Technology, Dongguan University of Technology, Dongguan, China}
\begin{abstract}
We investigate the stability of a circular electrodes-coated dielectric membrane under the combined action of an electric field and all-round in-plane tension. It is known that such a membrane is susceptible to the limiting point instability (also known as pull-in instability) which is widely believed to be a precursor to electric breakdown. However, there is experimental evidence showing that the limiting point instability may not necessarily be responsible for rapid thinning and electric breakdown. We explore the possibility that the latter is due to a new instability mechanism, namely localised axisymmetric necking. The bifurcation condition for axisymmetric necking is first derived and used to show that this instability may occur before the Treloar-Kearsley instability or the limiting point instability for a class of free energy functions. A weakly nonlinear analysis is then conducted and it is shown that the near-critical behavior is described by a fourth order nonlinear ODE with variable coefficients. This amplitude equation is solved using the finite difference method and it is demonstrated that a localised solution does indeed bifurcate from the homogeneous solution. Based on this analysis and what is already known for the purely mechanical case, we may deduce that the necking evolution follows the same three stages of initiation, growth and propagation as other similar localisation problems. The insight provided by the current study is expected to be relevant in assessing the integrity of dielectric elastomer actuators.
\end{abstract}
\begin{keyword}
Nonlinear electroelasticity\sep  dielectric membranes\sep localisation\sep stability\sep bifurcation
\end{keyword}
\end{frontmatter}
\section{Introduction}\label{introduction}
Dielectric elastomer actuators are believed to hold great potential in a wide range of applications such as  human-like robots, stretchable electronics,  and energy harvesting \citep{PK1998, PK00,CRK2008,CS2009, CB10, zllx2022}. It is known that such actuators are susceptible to a variety of instabilities \citep{PD06, ZW14}, and
before they can be deployed with confidence, a thorough understanding of their stability and buckling properties needs to be established. Thus, over the past two decades, much effort has been devoted to the understanding of the Hessian stability criterion \citep{zs2007, no2008, drs2008, DG10, xmkg2010, lzc2011, LH2012, ZW14, scd2019, lczy2021}, periodic wrinkling \citep{bg2011, rd2011,  do2014,  GC14, yzs2017b,  sb2018, do2019, gmz2019, scdd2020,brdo2020, xsc2021, badr2022,kjz2022}, \lq\lq two-phase" states \citep{PD06, ZH07, ZH08, ZK12, KZ12, HS12},  and the interplay between the limiting point instability and Treloar-Kearsley (TK) instability \citep{cy2021}. We refer to \cite{lm2020} for a comprehensive review of the relevant literature.

%cavitation \citep{yy2022},

The current study is concerned with a different kind of instability, namely necking,  that has received relatively less attention in the literature. Necking has traditionally been associated with ductile materials and plastic deformations, but in recent years it has been realised that elastic necking can occur in a wide range of soft materials under multiple fields; see, for instance, \cite{ntk2006}, \cite{mp2010}, \cite{zhao2012} and \cite{fjg2021}. The possibility of localised necking in a dielectric elastomer has previously been suggested by \cite{BL69} and analysed using an approximate model in a series of papers by \cite{pz2012}, \cite{zurlo2013}, \cite{DP13} and  \cite{zd2017}. The approximate model used in the latter papers is further discussed in \cite{fxd2018}.  For the case of uniaxial tension, localised necking was analysed by \cite{fdx2018} using analogies with the inflation problem associated with a rubber tube \citep{fpl2008}. It was shown that localised necking would initiate when the limiting point of nominal stress (as a function of stretch with fixed electric potential) or electric potential (as a function of electric displacement with fixed nominal stress) is reached. As in the inflation problem, the localised necking would evolve into a \lq\lq two-phase" deformation that has been observed experimentally by \cite{PD06}, and analysed by \cite{ZH07, ZH08, wg2019}.

Whereas the connection between the limiting-point instability and localised necking is now well understood in the case of uniaxial tension, this connection no longer exists in the case of equibiaxial tension, as demonstrated recently by \cite{wjf2022} and \cite{YF2022} for the purely mechanical case. For the case of equibiaxial tension,  the limiting-point behaviour may disappear at a large enough dead load, but some kind of snap-through behavior can still be observed that leads to pull-in failure \citep{HLF2012}. A likely scenario is that even if limiting-point instability does not exist, localised necking can still occur, and it is the axisymmetric necking that leads to a \lq\lq two-phase" deformation and possible pull-in failure. This scenario provides the major motivation for the current study. This paper may also be viewed as a sequel to our earlier paper, \cite{wjf2022}, where the axisymmetric necking was analysed in the purely mechanical context without an electric field. In that paper, the amplitude equation was left unsolved and it was not clear whether the equation did have a well-defined localised solution or not although fully numerical simulations seemed to have answered the question in the affirmative. In the current paper, we derive the corresponding results for the electroelastic case, and solve the amplitude equation to show that a localised solution does indeed bifurcate from the homogeneous solution.
%\cite{lk2013}: Giant voltage-induced deformation in dielectric elastomers near the verge of snap-through instability
%\cite{SG55}: earliest paper on electric breakdown
%Theory: \cite{do2005}, \cite{ML2005}, \cite{szg2008}, \cite{bdo2009}, \cite{Dorfmann2014a}
%Designing actuators that work on the verge of the voltage maximum \cite{ZW14}, \cite{yzs2017}
%The simplest prototypical actuator is a thin membrane coated with electrodes that is subjected to an electric potential across the thickness.

%\textcolor{red}{yu: index correspondence: $r\to 1$, $\theta\to 2$, $z\to 3$}

To set the context for our current study, consider a dielectric square membrane that is coated with electrodes and is subject to nominal tresses $S_1$ and $S_2$ in two mutually orthogonal directions within the membrane plane and a nominal electric field $E_3$ in the thickness direction (the $3$-direction). The associated stretches and nominal electric displacement are denoted by $\lambda_1, \lambda_2$ and $D_3$, respectively. In terms of the free energy function $\Omega(\lambda_1, \lambda_2, E_3)$, these quantities are related by \citep{do2005, zs2007}
\be
S_1=\frac{\paa \Omega}{\paa \lambda_1}, \;\;\;\; S_2=\frac{\paa \Omega}{\paa \lambda_2}, \;\;\;\;
D_3=- \frac{\paa \Omega}{\paa E_3}. \la{0.1} \en
Alternatively, defining $\Omega^*(\lambda_1, \lambda_2, D_3)=\Omega(\lambda_1, \lambda_2, E_3)+E_3 D_3$, we have
\be
S_1=\frac{\paa \Omega^*}{\paa \lambda_1}, \;\;\;\; S_2=\frac{\paa \Omega^*}{\paa \lambda_2}, \;\;\;\;
E_3=\frac{\paa \Omega^*}{\paa D_3}. \la{0.2} \en
The Hessian stability criterion states that the Hessian determinant defined by
\be
H=\left| \begin{array}{ccc} \vspace{0.2cm}\frac{\paa^2 \Omega^*}{\paa \lambda_1^2} &  \frac{\paa^2 \Omega^*}{\paa \lambda_1 \paa \lambda_2} &
 \frac{\paa^2 \Omega^*}{\paa \lambda_1 \paa D_3} \\ \vspace{0.2cm} \frac{\paa^2 \Omega^*}{\paa \lambda_2\paa \lambda_1 } &  \frac{\paa^2 \Omega^*}{\paa  \lambda_2^2} &
 \frac{\paa^2 \Omega^*}{\paa \lambda_2 \paa D_3} \\
 \frac{\paa^2 \Omega^*}{\paa D_3\paa \lambda_1 } &  \frac{\paa^2 \Omega^*}{\paa D_3 \paa  \lambda_2} &
 \frac{\paa^2 \Omega^*}{\paa D_3^2}
 \end{array} \right|=\left| \begin{array}{ccc} \vspace{0.2cm} \frac{\paa S_1}{\paa \lambda_1} &  \frac{\paa S_1}{\paa  \lambda_2} &
 \frac{\paa S_1}{ \paa D_3} \\\vspace{0.2cm} \frac{\paa S_2}{\paa \lambda_1 } &  \frac{\paa S_2}{\paa  \lambda_2} &
 \frac{\paa S_2}{ \paa D_3} \\
 \frac{\paa E_3}{\paa \lambda_1 } &  \frac{\paa E_3}{\paa  \lambda_2} &
 \frac{\paa E_3}{\paa D_3}
 \end{array} \right| \la{0.3} \en
should be positive definite for stability. Since $H=0$ is equivalent to $J(S_1, S_2, E_3)=0$ where the left-hand side denotes the Jacobian determinant of $S_1, S_2$ and $E_3$  in \rr{0.3}, marginal violation of the Hessian stability criterion means that the \lq\lq displacement" $(\lambda_1, \lambda_2, D_3)$ cannot uniquely be expressed in terms of the \lq\lq force" $(S_1, S_2, E_3)$. Evaluating the Jacobian determinant at equibiaxial stretching $\lambda_1=\lambda_2 \equiv \lambda$ where $S_1=S_2 \equiv S(\lambda,D_3)$, $E_3\equiv E(\lambda,D_3)$, $\paa S_1/\paa \lambda_2=\paa S_2/\paa \lambda_1$, $\paa S_1/\paa \lambda_1=\paa S_2/\paa \lambda_2$, ${\partial E_3}/{\partial\lambda_1}={\partial E_3}/{\partial\lambda_2}$, etc, we find that
\be
J(S_1, S_2, E_3)=\left(\frac{\paa S_1}{\paa \lambda_1}-\frac{\paa S_1}{\paa \lambda_2}\right) \left(\frac{\paa E}{\paa D_3} \frac{\paa S}{\paa \lambda}-  \frac{\paa S}{\paa D_3} \frac{\paa E}{\paa \lambda} \right), \la{0.4} \en
where all quantities are evaluated at $\lambda_1=\lambda_2 = \lambda$. The above expression may also be rewritten in two more revealing forms:
\be
J(S_1, S_2, E_3)= \Big(\frac{\paa S_1}{\paa \lambda_1}-\frac{\paa S_1}{\paa \lambda_2}\Big)\Big|_{E_3\; {\rm fixed}} \cdot \frac{\paa S(\lambda, D_3)}{\paa\lambda}\Big|_{E_3\; {\rm fixed}} \cdot \frac{\paa E_3}{\paa D_3}\Big|_{\lambda\; {\rm fixed}} , \la{0.5} \en
or
\be
J(S_1, S_2, E_3)= \Big(\frac{\paa S_1}{\paa \lambda_1}-\frac{\paa S_1}{\paa \lambda_2}\Big)\Big|_{D_3\; {\rm fixed}} \cdot \frac{\paa S(\lambda, D_3)}{\paa\lambda}\Big|_{D_3\; {\rm fixed}} \cdot \frac{\paa E_3}{\paa D_3} \Big|_{S \;{\rm fixed}}. \la{0.5a} \en

An application of  L$'$Hopital's rule gives the result
\be
\left.\left(\frac{\paa S_1}{\paa \lambda_1}-\frac{\paa S_1}{\paa \lambda_2}\right)\right|_{D_3\; {\rm fixed}}=\lim_{\lambda_2 \to \lambda_1} \left.\frac{S_2-S_1}{\lambda_2-\lambda_1}\right|_{D_3\; {\rm fixed}}. \la{0.5b} \en
It can also be shown that at equibiaxial stretching,
\be
\left.\left(\frac{\paa S_1}{\paa \lambda_1}-\frac{\paa S_1}{\paa \lambda_2}\right)\right|_{D_3\; {\rm fixed}}=\left.\left( \frac{\paa S_1}{\paa \lambda_1}-\frac{\paa S_1}{\paa \lambda_2}\right)\right|_{E_3\; {\rm fixed}}. \la{0.5c} \en
Thus, $H=J(S_1, S_2, E_3)=0$ is satisfied if any one of the following conditions is satisfied:
\be
 \lim_{\lambda_2 \to \lambda_1} \left.\frac{S_2-S_1}{\lambda_2-\lambda_1}\right|_{E_3\; {\rm fixed}} =0,\la{0.6} \en
\be
\left.\frac{\partial S}{\partial \lambda}\right|_{D_3 \;{\rm fixed}}=0, \;\;\;\; \left.\frac{\partial  S}{\partial \lambda}\right|_{E_3 \;{\rm fixed}}=0, \la{0.6a} \en
\be \left.\frac{\partial E_3}{\partial D_3}\right|_{\lambda\; {\rm fixed}} =0, \;\;\;\;
\left.\frac{\partial E_3}{\partial D_3}\right|_{S \;{\rm fixed}}=0. \la{0.6b} \en
The condition in \rr{0.6} obviously corresponds to the Treloar--Kearsley instability whereby unequal stretches occur at
equal nominal stresses \citep{og1985, ke1986, og1987}, whereas the other four conditions \rr{0.6a} and \rr{0.6b} correspond to the limiting points of $S$ and $E$, respectively. Also, it can be shown that \rr{0.6a}$_2$ and \rr{0.6b}$_2$ imply each other, and so we are left with four independent conditions. Only a subset of these four conditions can be satisfied depending on the material model adopted.
For instance, when the material is modelled as an ideal dielectric, the left-hand side of \rr{0.6b}$_1$ is always positive and \rr{0.6a}$_1$ is satisfied only after  \rr{0.6a}$_2$ is already satisfied. As a result, we are only left with two conditions: \rr{0.6b}$_2$ and \rr{0.6}. The former was the focus of study by  \cite{zs2007} and \cite{no2008}, whereas competition between the two conditions was studied by \cite{cy2021}.

It is commonly believed that when the Hessian stability criterion $H>0$ is violated, the dielectric membrane would thin down uniformly, leading eventually to electric breakdown or other types of failure (e.g. wrinkling). The result \rr{0.6} provides one counter-example to this common wisdom --  the TK instability may occur first before uniform thickness thinning takes place. In this paper, we explore another instability mechanism, namely localized axisymmetric necking whereby thickness thinning is localized near the origin and decays exponentially in the radial direction. Our preliminary investigations in Wang {\it et al} (2022) indicate that the condition for axisymmetric necking is not given by $H=0$ or the limiting point stability criterion although the necking condition in the case of plane-strain does correspond to the nominal stress reaching a limiting point \citep{fdx2018}. We observe that in the problem of localized bulging of an inflated hyperelastic tube, the bifurcation condition corresponds to the inflation pressure reaching a limiting point when the axial force is fixed or the axial force reaching a maximum when the pressure is fixed \citep{fi2015}.

The rest of this paper is divided into four sections as follows. In the next section we summarise the governing equations of electroelasticity and derive the incremental governing equations to the order that is required for the current analysis.  Sections 3 and 4 present the linear and weakly nonlinear analyses, respectively. The paper is concluded in Section 5 with a summary and some additional comments.

\section{Governing equations}\label{basic_equations}
\setcounter{equation}{0}
\subsection{Equations of nonlinear electroelasticity \label{electroelasticity}}
%%%%%%%%%%%%%%%%%%%%%%%%%%%%%%%%%%%%%%%%%%%%%%%%%%%%%%%%%%%%%%%%%%%%%%%%%%

%For more detailed background on nonlinear electroelasticity, we refer to the books by \citet{Landau1984}, \citet{Kovetz2000} and \citet{Dorfmann2014a}.

Consider a dielectric material that is free from volumetric free charges and mechanical body forces within the material and whose constitutive behavior is governed by the free energy density function $\Omega^*(F, \mathbf{D})$ or $\Omega(F, \mathbf{E})$ (=$\Omega^*(F, \mathbf{D})-\mathbf{D}\cdot \mathbf{E}$), where $F$ is the deformation gradient, $\mathbf{D}$ and $\mathbf{E}$ are the {\it nominal} electric displacement and electric field vectors, respectively. The nominal electric field, electric displacement, and the {\it total nominal stress tensor} $S$  satisfy the field equations
\begin{equation}
\label{field-eq}
\Curl\mathbf{E}=\mathbf{0},\quad \Div \mathbf{D}=0,\quad \Div S=\mathbf{0},
\end{equation}
where $\Curl$ and $\Div$ are the curl and divergence operators with respect to $\mathbf{X}$, the position vector in the undeformed configuration. The constitutive equations are either
\begin{equation}
S=\frac{\partial\Omega^*}{\partial F}-pF^{-1},\quad \mathbf{E}=\frac{\partial\Omega^*}{\partial\mathbf{D}},\label{law-1}
\end{equation}
or
\begin{equation}
S=\frac{\partial\Omega}{\partial F}-pF^{-1},\quad \mathbf{D}=-\frac{\partial\Omega}{\partial\mathbf{E}},\label{law-2}
\end{equation}
where we have assumed that the material is incompressible with $p$ denoting the Lagrangian multiplier enforcing the constraint of incompressibility $\det F=1$. See \cite{do2005} or \cite{zs2007} for further details.

It follows from \rr{field-eq}$_1$  that the electric field $\mathbf E$ can be specified in terms of an electrostatic potential $\Phi$:
\be \mathbf E=-\Grad \Phi. \la{pot} \en
We consider the case when the potential $\Phi$ is specified on the two surfaces of the membrane through the coating electrodes. As a result, the jump conditions at the interfaces between the membrane and surrounding medium need not be considered.

Following common practice, see, e.g., \citet{Dorfmann2014a}, we consider an energy function $\Omega( F, \mathbf E)$  that is additively decomposed as a purely mechanical contribution and a part associated with the electric field.
We further specialize to the case when the electric contribution is described by an isotropic constitutive formulation with constant permittivity $\epsilon$ (the so-called ideal dielectric). Thus, we have
\begin{equation}
\label{energy}
\Omega(F, \mathbf E) =W(I_1, I_2) -\frac{1}{2}  \epsilon  \,\mathbf E\cdot \mathbf C^{-1}  \mathbf E,
\end{equation}
where $I_1$ and $I_2$ are the two principal invariants of $FF^T$.
Correspondingly, in terms of the principal stretches the functions $\Omega$ and $\Omega^*$ in \rr{0.1} and \rr{0.2} take the specific forms
\be \Omega(\lambda_1, \lambda_2, E_3)=W(\lambda_1, \lambda_2)-\frac{1}{2} \vep E_3^2 (\lambda_1 \lambda_2)^2, \la{energy1} \en
\be \Omega^*(\lambda_1, \lambda_2, D_3)=W(\lambda_1, \lambda_2)+ \frac{1}{2\vep}  D_3^2 (\lambda_1 \lambda_2)^{-2}, \la{energy2} \en
where we have used the same symbols $\Omega$ and $W$ in \rr{energy} and \rr{energy1} (although the arguments are different) to avoid introducing extra notations. In the above equations, the incompressibility condition has been used to eliminate the principal stretch $\lambda_3$, and $W(\lambda_1, \lambda_2)$ is sometimes referred to as the {\it reduced strain energy function}.

For the above class of free energy functions, the left-hand side of \rr{0.6b}$_1$ is always positive  and we have
\be
\left.\frac{\partial S}{\partial \lambda}\right|_{E_3 \;{\rm fixed}}=\left.\frac{\partial S}{\partial\lambda}\right|_{D_3 \;{\rm fixed}}-8 \lambda^{2}  E_3^2 \vep. \la{limit10} \en
This means that \rr{0.6a}$_2$ is always satisfied before \rr{0.6a}$_1$ is satisfied. As a result, the conditions \rr{0.6b}$_1$ and \rr{0.6a}$_1$ can be neglected, and \rr{0.6}, \rr{0.6a}$_2$ can be solved explicitly (the condition \rr{0.6b}$_2$ is not independent as remarked earlier). Thus, we have the following two solutions
for the bifurcation values of $\vep E_3^2$:
\be
\left.\vep E_3^2\right|_{\rm TK}=\lambda^{-2} (W_{12}-W_{22}), \la{tk} \en
\be
\left.\vep E_3^2\right|_{\hbox{lim}}=\frac{1}{3} \lambda^{-2} (W_{12}+W_{22}), \la{lim1} \en
where the subscripts \lq\lq TK" and \lq\lq lim" signify associations with the TK and limiting-point instabilities, respectively,
and
\be W_{12}=\left.\frac{\paa^2 W}{\paa \lambda_1 \lambda_2}\right|_{\lambda_1= \lambda_2=\lambda}, \;\;\;\;
 W_{22}=\left.\frac{\paa^2 W}{\paa \lambda_2^2}\right|_{\lambda_1= \lambda_2=\lambda}. \la{aux} \en
\subsection{Incremental formulation}
%%%%%%%%%%%%%%%%%%%%%%%%%%%%%%%%%%%%%%%%%%%%%%%%%%%%%%%%%%%%%%%%%%%%%%%%%%

In this section we derive the equations governing incremental deformations up to and including quadratic terms. For the linear version, see \cite{Dorfmann2010}.

We denote the undeformed, uniformly stretched, and bifurcated configurations of the membrane by $B_0$, $B_e$ and $B_t$, and the position vectors of a representative material particle in the three configurations by ${\bm X}$, ${\bm x}$ and $\tilde{\bm x}$, respectively. We use $F$, ${\bm E}$, ${\bm D}$ and $S$  to denote the deformation gradient, the nominal electric field, nominal electric displacement and total nominal stress associated with the deformation $B_0$ $\to$ $B_t$. Their counterparts associated with the deformation $B_0$ $\to$ $B_e$ are denoted by $\bar{F}$, $\bar{\bm E}$, $\bar{\bm D}$ and $\bar{S}$. We define the incremental fields $\eta$, ${\bm e}$, ${\bm d}$, and $\chi$ through
%We write $\tilde{\bm x}={\bm x}+{\bm u}({\bf x})$ with ${\bm u}$ signifying the incremental displacement from $B_e$ to $B_t$. With the deformation gradients from $B_0$ to $B_e$, $B_0$ to $B_t$, and $B_e$ to $B_t$  denoted by $\bar{F}, F$ and $I+\eta$, respectively, where $\eta={\rm grad}\, {\bm u}$, it follows that
\be F=(I+\eta) \bar{F}, \;\;\;\; {\bm E}=\bar{\bm E}+\bar{F}^T {\bm e}, \la{aug1a} \en \be {\bm D}=\bar{\bm D}+ \bar{J} \bar{F}^{-1} {\bm d}, \;\;\;\; S=\bar{S}+\bar{J} \bar{F}^{-1} \chi^T. \la{aug1} \en
The determinant $\bar{J}$ ($={\rm det}\,\bar{F}$) is unity but is kept in the above expressions to maintain the generality of the formulae.
With ${\bm u}({\bm x})$ denoting the incremental displacement from $B_e$ to $B_t$, we have  $\eta={\rm grad}\,{\bm u}$. From the governing equations \rr{field-eq} that apply to both the barred and unbarred fields, we obtain the incremental governing equations
\be {\rm curl}\, {\bm e}={\bm 0}, \;\;\;\; {\rm div}\, {\bm d}=0, \;\;\;\; {\rm div}\, \chi^T={\bm 0}, \la{aug0} \en
where div and curl are evaluated with respect to the position vector ${\bm x}$.

We now proceed to derive the incremental forms of the constitutive equations \rr{law-2}. We first expand
$\paa \Omega/\paa F_{iA}$ around $F=\bar{F},\; {\bm E}=\bar{\bm E}$ to obtain
%$$
%\frac{\paa \Omega}{\paa F_{iA}}=\left.\frac{\paa \Omega}{\paa F_{iA}}\right|_{\bar{F}}+\left.\frac{\paa^2 \Omega}{\paa F_{iA}\paa F_{kB}}\right|_{\bar{F}} \eta_{kj} \bar{F}_{jB}+
%\left.\frac{\paa^2 \Omega}{\paa F_{iA}\paa E_{B}}\right|_{\bar{F}} e_k \bar{F}_{kB} $$
%$$
%+\frac{1}{2} \left.\frac{\paa^3 \Omega}{\paa F_{iA}\paa F_{kB}\paa F_{mC}}\right|_{\bar{F}} \eta_{kj} \bar{F}_{jB}\eta_{mn} \bar{F}_{nC}+
%\left.\frac{\paa^3 \Omega}{\paa F_{iA}\paa F_{kB}\paa E_{C}}\right|_{\bar{F}} \eta_{kj} \bar{F}_{jB} e_n \bar{F}_{nC}$$ \be +\frac{1}{2} \left.\frac{\paa^3 \Omega}{\paa F_{iA}\paa E_B \paa E_C}\right|_{\bar{F}} e_j  \bar{F}_{jB} e_k  \bar{F}_{kC}. \la{aug2} \en
%It then follows that
%$$
%\frac{\paa \Omega}{\paa F_{iA}}\bar{F}_{lA}=\bar{F}_{lA}  \left.\frac{\paa \Omega}{\paa F_{iA}}\right|_{\bar{F}}+\bar{F}_{lA} \bar{F}_{jB} \left.\frac{\paa^2 \Omega}{\paa F_{iA}\paa F_{kB}}\right|_{\bar{F}} \eta_{kj}+
%\bar{F}_{lA} \bar{F}_{kB} \left.\frac{\paa^2 \Omega}{\paa F_{iA}\paa E_{B}}\right|_{\bar{F}} e_k   $$
%$$
%+\frac{1}{2} \bar{F}_{lA} \bar{F}_{jB} \bar{F}_{nC} \left.\frac{\paa^3 \Omega}{\paa F_{iA}\paa F_{kB} \paa F_{mC}}\right|_{\bar{F}} \eta_{kj}  \eta_{mn}  +
%\bar{F}_{lA} \bar{F}_{jB} \bar{F}_{nC} \left.\frac{\paa^3 \Omega}{\paa F_{iA}\paa F_{kB}\paa E_{C}}\right|_{\bar{F}} \eta_{kj}   e_n  $$ \be +\frac{1}{2} \bar{F}_{lA} \bar{F}_{jB} \bar{F}_{kC} \left.\frac{\paa^3 \Omega}{\paa F_{iA}\paa E_B \paa E_C}\right|_{\bar{F}} e_j    e_k. \la{aug2a} \en
%These results enable us to evaluate
$$
 \left(\bar{J}^{-1} \bar{F}  \frac{\paa \Omega}{\paa F}\right)_{li}=\bar{J}^{-1} \bar{F}_{lA} \frac{\paa \Omega}{\paa F_{iA}}=\bar{J}^{-1} \bar{F}_{lA}  \left.\frac{\paa \Omega}{\paa F_{iA}}\right|_{\bar{F}}+\mathcal{A}_{lijk}^{(1)} \eta_{kj}+ \mathbb{A}^{(1)}_{li|k} e_k   $$
\be
+\frac{1}{2}\mathcal{A}^{(2)}_{lijknm} \eta_{kj}  \eta_{mn} +\mathbb{A}^{(2)}_{lijk|n} \eta_{kj}   e_n  +\frac{1}{2} \mathbb{A}^{(3)}_{li|jk}  e_j    e_k, \la{aug2b} \en
where
\be
\label{moduli-1}
\mathcal{A}^{(1)}_{lijk}=\bar{J}^{-1}\bar{F}_{l A} \bar{F}_{j B} \left.\frac{\partial^2 \Omega}{\partial F_{i A} \partial F_{k B}}\right|_{\bar{F}}, \quad
\mathcal{A}^{(2)}_{lijknm}=\bar{J}^{-1}\bar{F}_{l A} \bar{F}_{j B} \bar{F}_{n C} \left.\frac{\partial^2 \Omega}{\partial F_{i A} \partial F_{k B} \paa F_{mC}}\right|_{\bar{F}},
\en
\be
\mathbb{A}^{(1)}_{li|k}=\bar{J}^{-1} \bar{F}_{lA}\bar{F}_{kB}\left.\frac{\partial ^2 \Omega}{\partial F_{i A} \partial E_{ B}}\right|_{\bar{F}}, \quad
\mathbb{A}^{(2)}_{lijk|n}=\bar{J}^{-1} \bar{F}_{lA} \bar{F}_{jB} \bar{F}_{nC} \left.\frac{\paa^3 \Omega}{\paa F_{iA}\paa F_{kB}\paa E_{C}}\right|_{\bar{F}},
\label{moduli-2}
\end{equation}
\be
\mathbb{A}^{(3)}_{li|jk}=\bar{J}^{-1} \bar{F}_{lA} \bar{F}_{jB} \bar{F}_{kC} \left.\frac{\paa^3 \Omega}{\paa F_{iA}\paa E_B \paa E_C}\right|_{\bar{F}}. \la{moduli-3} \en
We also have
\be
p \bar{ F} { F}^{-1}   =(\bar{p}+p^*)  (I+\eta)^{-1}  =\bar{p} (
{ I} - \eta + \eta^{2} )+p^* (
{ I} - \eta )  + \cdots \la{aug3} \en
Thus, it follows from \rr{law-2}$_1$ and \rr{aug1a}$_2$  that
$$
(\chi^T)_{li}=\mathcal{A}^{(1)}_{lijk} \eta_{kj}+ \mathbb{A}^{(1)}_{li|k} e_k
+\frac{1}{2}\mathcal{A}^{(2)}_{lijknm} \eta_{kj}  \eta_{mn} +\mathbb{A}^{(2)}_{lijk|n} \eta_{kj}   e_n  +\frac{1}{2} \mathbb{A}^{(3)}_{li|jk}  e_je_k $$
\be
+  \bar{p}   (\eta_{li} - \eta_{lk} \eta_{ki})
 -  p^{*}   (\delta_{li} - \eta_{li} )+\cdots.  \la{aug4} \en
For the electric displacement, we can similarly obtain
%$$
%\frac{\paa \Omega}{\paa E_M}=\left. \frac{\paa \Omega}{\paa E_M} \right|_{\bar{F}}+\left. \frac{\paa^2 \Omega}{\paa F_{iA}\paa E_M} \right|_{\bar{F}} \eta_{im} \bar{F}_{mA}+
%\left. \frac{\paa^2 \Omega}{\paa E_A \paa E_M} \right|_{\bar{F}}  e_j \bar{F}_{jA} $$
%$$
%+\frac{1}{2} \left. \frac{\paa^3 \Omega}{\paa F_{iA} \paa F_{kB} \paa E_M} \right|_{\bar{F}} \eta_{im} \bar{F}_{mA}\eta_{kn} \bar{F}_{nB}+
%\left. \frac{\paa^3 \Omega}{\paa F_{iA} \paa E_C \paa E_M} \right|_{\bar{F}} \eta_{im} \bar{F}_{mA} e_n \bar{F}_{nC} $$ \be +
%\frac{1}{2} \left. \frac{\paa^3 \Omega}{\paa E_A \paa E_C \paa E_M} \right|_{\bar{F}} e_i \bar{F}_{iA} e_n \bar{F}_{nC}+\cdots, \la{aug5} \en
$$
\bar{J}^{-1} \bar{F}_{lM} \frac{\paa \Omega}{\paa E_M}=\bar{J}^{-1} \bar{F}_{lM} \left. \frac{\paa \Omega}{\paa E_M} \right|_{\bar{F}}+\bar{J}^{-1} \bar{F}_{lM}\bar{F}_{mA} \left. \frac{\paa^2 \Omega}{\paa F_{iA}\paa E_M} \right|_{\bar{F}} \eta_{im} +\bar{J}^{-1} \bar{F}_{lM} \bar{F}_{jA}
\left. \frac{\paa^2 \Omega}{\paa E_A \paa E_M} \right|_{\bar{F}}  e_j  $$
$$
+\frac{1}{2} \bar{J}^{-1} \bar{F}_{lM} \bar{F}_{mA}\bar{F}_{nB}   \left. \frac{\paa^3 \Omega}{\paa F_{iA} \paa F_{kB} \paa E_M} \right|_{\bar{F}} \eta_{im}\eta_{kn} +
\bar{J}^{-1} \bar{F}_{lM} \bar{F}_{mA} \bar{F}_{nC} \left. \frac{\paa^3 \Omega}{\paa F_{iA} \paa E_C \paa E_M} \right|_{\bar{F}} \eta_{im}  e_n  $$ \be +
\frac{1}{2} \bar{J}^{-1} \bar{F}_{lM}\bar{F}_{iA} \bar{F}_{nC} \left. \frac{\paa^3 \Omega}{\paa E_A \paa E_C \paa E_M} \right|_{\bar{F}} e_i  e_n +\cdots. \la{aug5} \en
It then follows from \rr{aug1}$_1$ and \rr{law-2}$_2$ that
%Thus, if ${\bm d}=\bar{J}^{-1} \bar{F} ({\bm D}-\bar{\bm D}) $, then
$$ d_l=\bar{J}^{-1} \bar{F}_{lM} (-\frac{\paa \Omega}{\paa E_M}+\left.\frac{\paa \Omega}{\paa E_M}\right|_{\bar{F}})=-\mathbb{A}^{(1)}_{mi|l} \eta_{im}-\mathsf{A}^{(1)}_{jl} e_j $$
\be
-\frac{1}{2} \mathbb{A}^{(2)}_{mink|l}  \eta_{im}\eta_{kn}-\mathbb{A}^{(3)}_{mi|nl}  \eta_{im}e_n- \frac{1}{2} \mathsf{A}^{(2)}_{iln} e_i e_n+\cdots, \la{aug6} \en
where
\begin{equation}
\mathsf{A}^{(1)}_{jl}= \bar{J}^{-1} \bar{F}_{j A}  \bar{F}_{l B} \left.\frac{\partial^2 \Omega}{\partial E_{ A} \partial E_{ B}}\right|_{\bar{F}},\;\;\;\;
\mathsf{A}^{(2)}_{iln}= \bar{J}^{-1} \bar{F}_{iA} \bar{F}_{lM} \bar{F}_{nC} \left. \frac{\paa^3 \Omega}{\paa E_A \paa E_M \paa E_C } \right|_{\bar{F}}.
\label{moduli-33}
\end{equation}
Finally, it follows from the incompressibility conditions ${\rm det}\,\bar{ F}=1 $ and ${\rm det}\, { F}=1 $ that
\be
I_\eta+II_\eta+III_\eta=0, \la{incom2} \en
where the three terms denote the three principal invariants of $\eta$, respectively. This is the incremental incompressibility condition and its linear form is simply ${\rm tr}\, \eta ={\rm div}\, {\bm u}=0$.

%These definitions imply the symmetries
%\begin{equation}
%\label{A-kappa-symmetries}
%\mathcal{A}_{ijlk}=\mathcal{A}_{lkij},\quad \mathbb{A}_{ij|l}=\mathbb{A}_{ji|l}, \quad \mathsf{A}_{ij}=\mathsf{A}_{ji}.
%\end{equation}
%For our current problem, the principal directions of deformation coincide with the three coordinate axes  and we shall view $\Omega$ as a function of $\lambda_1, \lambda_3$ and $E_2$ and compute the moduli $\mathcal{A}^{(1)}_{lijk}$ according to the formulae
%\begin{eqnarray}
%{\cal A}^{(1)}_{iijj}&=&\lambda_i\lambda_j\Omega_{ij},\nonumber \\
%{\cal A}^{(1)}_{ijkl}&=&\frac{\lambda_i^2}{\lambda_i^2-\lambda_j^2}
%(\lambda_i\Omega_i-\lambda_j\Omega_j)\delta_{ik}\delta_{jl} \nonumber \\
%& &  +\frac{\lambda_i\lambda_j}{\lambda_i^2-\lambda_j^2}
%(\lambda_j\Omega_i-\lambda_i\Omega_j)\delta_{il}\delta_{jk}, \;\;\;\;\;
%i\ne j,\la{moduli}
%\end{eqnarray}
%where $\Omega_i=\paa \Omega/\paa\lambda_i,\,
%\Omega_{ij}=\paa^2\Omega/\paa\lambda_i\paa\lambda_j$ etc, and the principal stretches are understood to correspond to the
%primary deformation $B_0 \to B_e$. It is straightforward to verify that
%\be {\cal A}_{1212}-{\cal A}_{1221}=\lambda_1 \Omega_1,\;\;\;\; {\cal A}_{2121}-{\cal A}_{2112}=\lambda_2 \Omega_2. \la{connect} \en
%
%

%The incremental equilibrium equations are then given by
%\be
%{\rm div}\, \chi^T =0, \;\;\;\; {\rm div}\, {\bm d}=0. \la{aug6} \en
The governing equation \rr{aug0}$_{1}$ can be satisfied automatically by writing
${\bm e}={\rm grad}\, \psi$ where the scalar function $\psi$ replaces ${\bm e}$ as one of the new independent variables. The remaining governing equations \rr{aug0}$_{2,3}$  are to be solved subjected to the boundary conditions
\be \chi_{33}=0,\;\;\;\; \chi_{31}=0, \;\;\;\; \psi=0 \;\;\;\;{\rm on}\;\;  z= \pm h/2. \la{incre5a} \en
We take $h=1$ in the remaining analysis, which is equivalent to using $h$ as the length unit.

\section{Linear analysis}
\setcounter{equation}{0}
We now consider an axisymmetric perturbation represented by
\be {\bm u} =u(r, z) {\bm e}_r+v(r, z) {\bm e}_z, \;\;\;\; \psi=\psi(r, z), \la{incrr} \en
where $r$ and $\theta$ are the cylindrical coordinates for ${\bm x}$, ${\bm e}_r$ and ${\bm e}_z$  are the unit basis vectors, and $u$ and $v$ are the associated displacement components.
The tensor $\eta$ ($={\rm grad}\, {\bm u}$) now takes the form
\be
\eta =u_r {\bm e}_r \otimes {\bm e}_r+u_z {\bm e}_r \otimes {\bm e}_z+ \frac{u}{r} {\bm e}_\theta \otimes {\bm e}_\theta+v_r  {\bm e}_z \otimes {\bm e}_r + v_z  {\bm e}_z \otimes {\bm e}_z, \la{eta} \en
where $u_r=\paa u/\paa r$, $u_z=\paa u/\paa z$, etc.

For the current axisymmetric problem, the two components of the equilibrium equation ${\rm div}\, \chi^T =0$ that are not satisfied automatically are
\be
\chi_{1j,j}+\frac{1}{r} (\chi_{11}-\chi_{22})=0, \;\;\;\; \chi_{3j,j}+\frac{1}{r} \chi_{31}=0, \la{incre1} \en
where $(1, 2, 3)$ corresponds to $(r, \theta, z)$.
The linearization of the incompressibility condition \rr{incom2}, namely ${\rm div}\, {\bm u}=0$, may be written in the form
\be \frac{\paa\, (r u)}{\paa r}+\frac{ \paa\, (r v)}{\paa z}=0, \la{3.1} \en
which can be satisfied automatically by introducing a \lq stream function\rq $\,\phi(r, z)$ such that
\be u=\frac{1}{r} \phi_z, \;\;\;\; v=-\frac{1}{r} \phi_r, \la{3.2} \en
where as in \rr{eta} a subscript signifies differentiation (e.g. $\phi_z=\paa \phi/\paa z$).
The non-zero stress components are given by
\begin{eqnarray}
\chi_{11}&=& \mathcal{A}^{(1)}_{1122} \frac{u}{r} + \mathcal{A}^{(1)}_{1133} v_z+(\mathcal{A}^{(1)}_{1111}+\bar{p}) u_r-p^*, \noindent \\
 \chi_{22}&=&(\mathcal{A}^{(1)}_{2222}+\bar{p}) \frac{u}{r} +\mathcal{A}^{(1)}_{2233} v_z+\mathcal{A}^{(1)}_{1122} u_r-p^*, \noindent \\
  \chi_{33}&=& \mathcal{A}^{(1)}_{2233} \frac{u}{r} +(\mathcal{A}^{(1)}_{3333}+\bar{p}) v_z+\mathcal{A}^{(1)}_{1133} u_r-p^*-2 E_3 \vep \lambda^2 \psi_z, \noindent \\
  \chi_{13}&=& \mathcal{A}^{(1)}_{3131} u_z+(\mathcal{A}^{(1)}_{3113}+\bar{p}) v_r- E_3 \vep \lambda^2 \psi_r, \noindent\\
  \chi_{31}&=& \mathcal{A}^{(1)}_{1313} v_r+(\mathcal{A}^{(1)}_{1331}+\bar{p}) u_z - E_3 \vep \lambda^2 \psi_r, \noindent
  \end{eqnarray}
whereas the linearisation of \rr{aug6} is given by
\be
d_1= - E_3 \vep \lambda^2 (u_z+v_r)-E_3 \psi_r, \;\;\;\;  d_2=0, \;\;\;\; d_3=-2 E_3 \vep \lambda^2 v_z-E_3 \psi_z.  \la{lineard} \en
On substituting these expressions together with \rr{3.2} into \rr{incre1} and then eliminating $p^*$ by cross-differentiation, we obtain
$$
\alpha \left( \phi_{rrrr}-\frac{2}{r} \phi_{rrr}+\frac{3}{r^2} \phi_{rr}-\frac{3}{r^3} \phi_{r}  \right)+2 \beta \left( \phi_{rrzz}-\frac{1}{r} \phi_{rzz} \right)+\gamma \phi_{zzzz}
$$
\be +E_3 \vep \lambda^2 \left( r \psi_{rrr}+\psi_{rr}-\frac{1}{r} \psi_r+r \psi_{rzz} \right)
=0, \la{3.4} \en
where
\be \alpha=\mathcal{A}^{(1)}_{2323}, \;\;\;\; 2 \beta=\mathcal{A}^{(1)}_{2222}+\mathcal{A}^{(1)}_{3333}-2 \mathcal{A}^{(1)}_{2233}-2 \mathcal{A}^{(1)}_{2332}, \;\;\;\; \gamma=\mathcal{A}^{(1)}_{3232}. \la{3.5} \en
A second equation for $\phi$ and $\psi$ is obtained by substituting \rr{lineard} into \rr{aug0}$_2$:
\be
\psi_{zz}+\frac{1}{r} \psi_r+\psi_{rr}-   E_3 \lambda^2 \frac{1}{r^3} \left(r^2 \phi_{rrr}-r \phi_{rr}+r^2 \phi_{rzz}+\phi_{r} \right)=0. \la{eqnpsi}
\en
Equation \rr{3.4}and \rr{eqnpsi} admit a \lq\lq normal mode" buckling/wrinkling solution of the form
\be \phi(r, z)=r J_1(k r) S(k z), \;\;\;\; \psi(r, z)=J_0(k r) K(k z),  \la{add1} \en
where $k$ is a constant playing the role of wavenumber, $J_0(x)$ and $J_1(x)$ are Bessel's functions of the first kind, and the other functions $S(k z)$ and $K(k z)$ are to be determined. %This step is similar to looking for periodic buckling solutions proportional to ${\rm e}^{\ii k x_1}$ in rectangular coordinates. In fact, $J_1(x)$ is oscillatory although it also decays like $1/\sqrt{x}$ for large values of $x$ due to geometric spreading.

On substituting \rr{add1} into  \rr{3.4} and  \rr{eqnpsi} and simplifying by making use of the identity
$$ J_\nu (x)= \frac{2 (\nu -1)}{x} J_{\nu-1}(x)-J_{\nu-2}(x), $$
the $J_1(k r)$ and $J_0(k r)$ can be cancelled in the resulting equations and we obtain two ordinary differential equations:
\be \gamma S^{(4)}(k z)-2 \beta S''(k z)+\alpha S (k z)+ k^{-1} E_3 \vep \lambda^2 (K(k z)-K''(k z) )=0, \la{aug10} \en
and
\be
 \left\{K''(k z)-E_3 k \lambda^2 S''(k z)\right\}-\left\{ K(k z)-E_3 k \lambda^2 S(k z)\right\} =0. \la{aug11} \en
The last equation can be integrated straightaway to yield
\be
K(k z)=E_3 k \lambda^2 S(k z)+c_5\, {\rm sinh}(k z)+c_6\, {\rm cosh}(k z), \la{solK} \en
where $c_5$ and $c_6$ are constants. Equation \rr{aug10} then reduces to
\be \gamma S^{(4)}(k z)-2 \beta^* S''(k z)+\alpha^* S (k z)=0, \la{add2} \en
where
\be \alpha^*=\alpha+  E_3^2 \vep \lambda^4, \;\;\;\; \beta^*=\beta+\frac{1}{2} E_3^2 \vep \lambda^4. \la{abc} \en
The general solution of \rr{add2} may be written in the form
\be S(k z) = c_1 \sinh \frac{k}{\sqrt{\zeta_1} } z+
c_2 \sinh \frac{k}{\sqrt{\zeta_2} } z+c_3 \cosh \frac{k}{\sqrt{\zeta_1} } z +c_4 \cosh \frac{k}{\sqrt{\zeta_2} } z, \la{3.15} \en
where $c_1, c_2, c_3, c_4$ are disposable constants, and
 \be \zeta_1=\frac{1}{\alpha^*} (\beta^*-\sqrt{\beta^{*2}-\alpha^* \gamma}), \;\;\;\;
 \zeta_2=\frac{1}{\alpha^*} (\beta^*+\sqrt{\beta^{*2}-\alpha^* \gamma}). \la{3.8} \en
The boundary conditions \rr{incre5a} take the form
\be
 {\cal A}^{(1)}_{3131}  u_z +({\cal A}^{(1)}_{3113}+\bar{p}) v_r-E_3 \vep \lambda^2 \psi_r=0, \;\;\;\; {\rm on}\;\; z=\pm 1/2, \la{3.16} \en
\be
 {\cal A}^{(1)}_{2233} \frac{u}{r} +({\cal A}^{(1)}_{3333}+\bar{p}) v_z+{\cal A}^{(1)}_{1133} u_r-p^*-2 E_3 \vep \lambda^2 \psi_z=0,\;\;\;\; {\rm on}\;\; z=\pm 1/2, \la{3.17} \en
 \be \psi=0, \;\;\;\; {\rm on}\;\; z=\pm 1/2. \la{3.17a} \en
The  $p^*$ in \rr{3.17} can be eliminated by first differentiating \rr{3.17} with respect to $r$ and then using \rr{incre1}$_1$ to eliminate $p^*_r$. This gives
 $$ \frac{1}{r^2} ({\cal A}^{(1)}_{2233}-{\cal A}^{(1)}_{2222}-\bar{p}) \left(r^2 u_{rr}+r u_r-u\right)-{\cal A}^{(1)}_{3232} u_{zz} $$
\be  +({\cal A}^{(1)}_{3333}-{\cal A}^{(1)}_{2332}-{\cal A}^{(1)}_{2233}) v_{rz}-E_3 \vep \lambda^2 \psi_{rz}=0, \;\;\;\; {\rm on}\;\; z=\pm 1/2. \la{3.18} \en
On substituting \rr{add1}, \rr{solK} and \rr{3.15}  into the six boundary conditions \rr{3.16}, \rr{3.17a} and \rr{3.18}, we obtain six algebraic equations. Due to the symmetry of the membrane geometry and external loads with respect to the mid-plane $z=0$, this system of equations admits two types of solutions corresponding to flexural and extensional modes, respectively.
%By simple addition and subtraction, the six equations can be decoupled into two equations for $c_1$, $c_2$ and $c_5$ and another three equations for $c_3$, $c_4$ and $c_6$. Since the determinants of the two coefficient matrices cannot vanish simultaneously,  when the equations for $(c_1, c_2, c_5)$ have non-trivial solutions, $(c_3, c_4, c_6)$ must vanish. In this case, $\phi$, and hence the vertical displacement $v$, are odd functions of $z$. The associated modes are thus extensional. On the other hand, when the equations for $(c_3, c_4, c_6)$  have non-trivial solutions, $(c_1, c_2, c_5)$ must vanish, and the associated modes are flexural.
The bifurcation condition for the extensional modes is what we shall focus on and is given by
$$
d_1 \tanh \left(\frac{k  }{2}\right) \tanh \left(\frac{k  }{2
   \sqrt{\zeta_1}}\right)-d_2 \tanh \left(\frac{k  }{2}\right) \tanh \left(\frac{k  }{2
   \sqrt{\zeta_2}}\right)$$ \be +d_3 \tanh \left(\frac{k  }{2
   \sqrt{\zeta_1}}\right) \tanh \left(\frac{k  }{2
   \sqrt{\zeta_2}}\right) =0, \la{disp} \en
   where
\begin{eqnarray}
d_1&=&\sqrt{\zeta_1} (1+\zeta_1)(\zeta_2 (2 \beta^*+\gamma)-\gamma), \nonumber \\
d_2&=&\sqrt{\zeta_2} (1+\zeta_2)(\zeta_1 (2 \beta^*+\gamma)-\gamma), \la{d2} \\
   d_3&=&\vep E_3^2 \lambda ^4 \sqrt{\zeta_1\zeta_2} (\zeta_1-\zeta_2). \nonumber \end{eqnarray}
Expanding \rr{disp} to order $k^2$, we obtain
%$$ \gamma ( \beta+ \gamma)+ \frac{k^2}{48} \left\{2 \gamma  \alpha _s+6
%   \gamma  \beta _s+4 \beta _s^2 \right. $$ $$ \hspace{2cm} \left. - E_3^2 \vep \lambda^4  (3 \gamma+2 \beta_s) \right\}+O(k^4)=0,$$
%   or equivalently,
\be \gamma ( \beta+ \gamma)+ \frac{k^2}{24} \gamma (\alpha-\gamma)+O(k^4)=0, \la{asy5} \en
where we have used \rr{3.8} to eliminate $\zeta_1$ and $\zeta_2$. Note that the coefficient of $k^2$ in the above asymptotic expression is not unique: we can add an arbitrary multiple of $\gamma ( \beta+ \gamma)$ to it without changing the asymptotic order of the second term since the latter expression is of order $k^2$.

\begin{figure}[ht]
\begin{center}
\begin{tabular}{ccc}
\includegraphics[width=.45\textwidth]{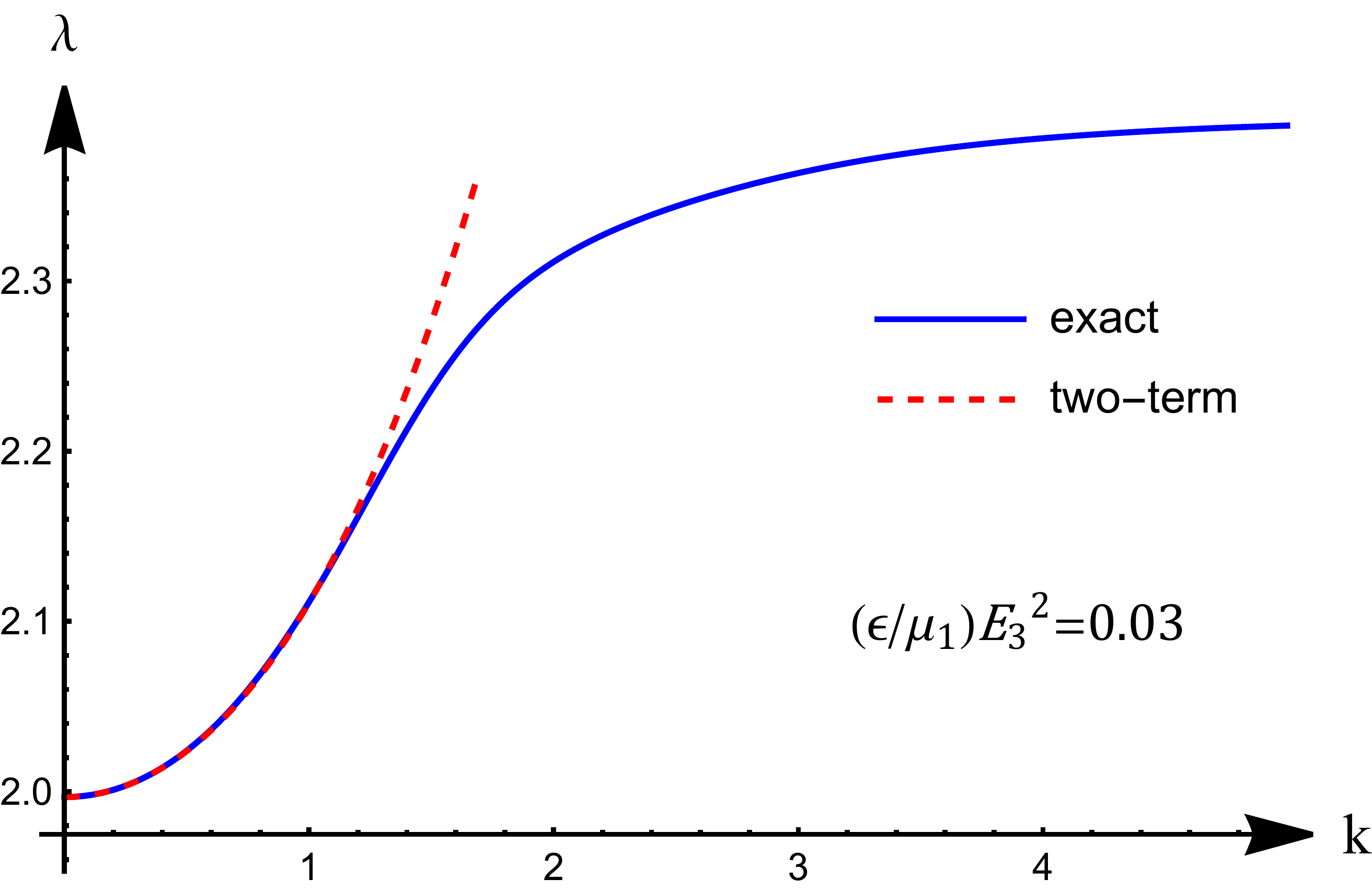}  & & \includegraphics[width=.45\textwidth]{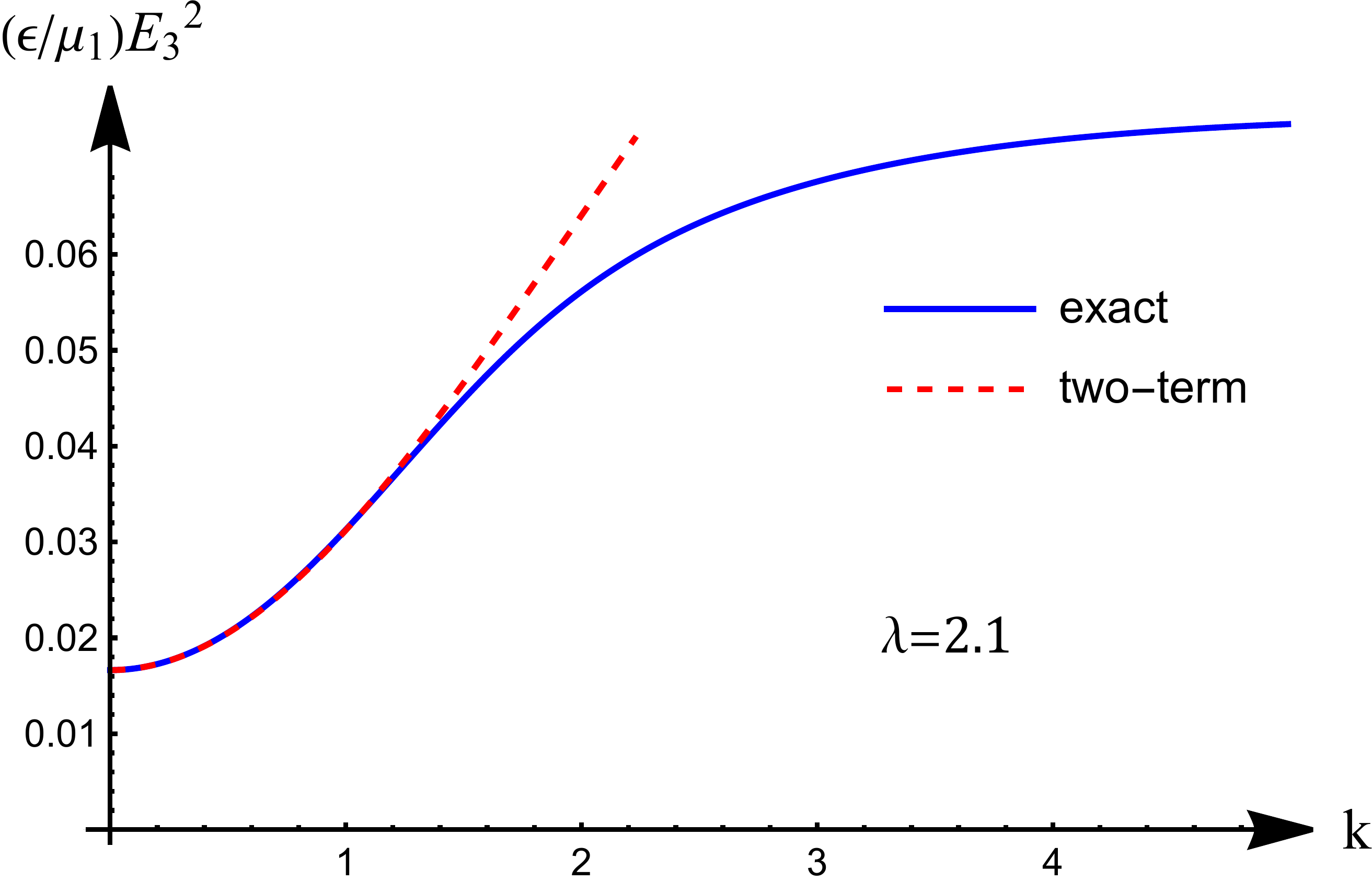}  \\ \\
(a) & & (b)
\end{tabular}
\caption{Bifurcation condition \rr{disp} for periodic and symmetric modes, and its two-term approximation \rr{asy5} in the small wavenumber limit.}
\label{fig0}
\end{center}
\end{figure}

As an illustrative example, consider the following two-term Ogden strain-energy function:
 \be
W=\frac{2\mu_1}{m_1^2} (\lambda_1^{m_1}+\lambda_2^{m_1}+\lambda_3^{m_1} -3 )+\frac{2\mu_2}{{m_2}^2} (\lambda_1^{m_2}+\lambda_2^{m_2}+\lambda_3^{m_2} -3 ), \la{energy_2terms}
\en
with $m_1=1/2, \; m_2=4, \mu_2=\mu_1/80$. Fig.~\ref{fig0} displays the bifurcation condition \rr{disp} and its two-term approximation \rr{asy5} in the small wavenumber limit. It is seen that the the minimum of $\lambda$ is attained at $k=0$ in the case of fixed $E_3$ and the minimum of $E_3$ is also attained at $k=0$ in the case of fixed $\lambda$.
Based on the discussion in \cite{fu2001}, we may postulate that the bifurcation condition for localized necking can be obtained by setting the leading order term in \rr{asy5} to zero, that is $ \beta +\gamma=0$ since $\gamma>0$, or equivalently,
\be \mathcal{A}^{(1)}_{2222}+\mathcal{A}^{(1)}_{3333}+2 \mathcal{A}^{(1)}_{3232}-2 \mathcal{A}^{(1)}_{2332}-2 \mathcal{A}^{(1)}_{2233}=0. \la{bif} \en
It can be shown that this condition is equivalent to
\be
\left.\frac{\paa S_1}{\paa \lambda_1}\right|_{\lambda_1=\lambda_2=\lambda}=0, \la{equivalent} \en
where $S_1$ has the same meaning as in Section 1. Corresponding to the free energy function \rr{energy1}, this equation can be solved explicitly to give
\be
(\vep E_3^2)_{\rm necking}=\lambda^{-2} W_{11}, \la{necking-condition} \en
where
\be W_{11}=\left.\frac{\paa^2 W}{\paa \lambda_1^2}\right|_{\lambda_1= \lambda_2=\lambda}. \la{w33} \en
The bifurcation condition may be compared with the conditions \rr{tk} and \rr{lim1} for the TK and limiting point instabilities.

\begin{figure}[ht]
\begin{center}
\begin{tabular}{ccc}
\includegraphics[width=.45\textwidth]{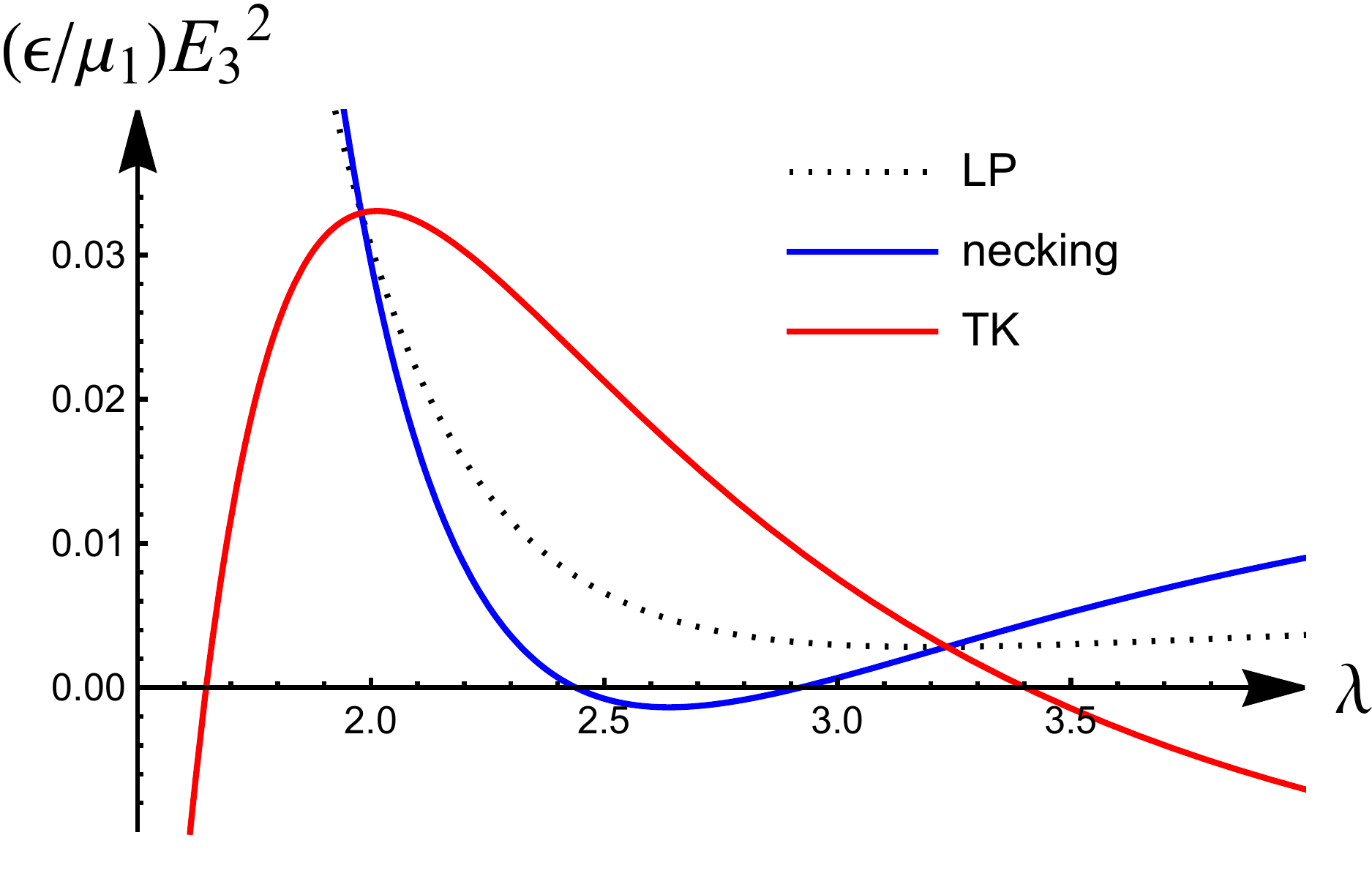} & &\includegraphics[width=.45\textwidth]{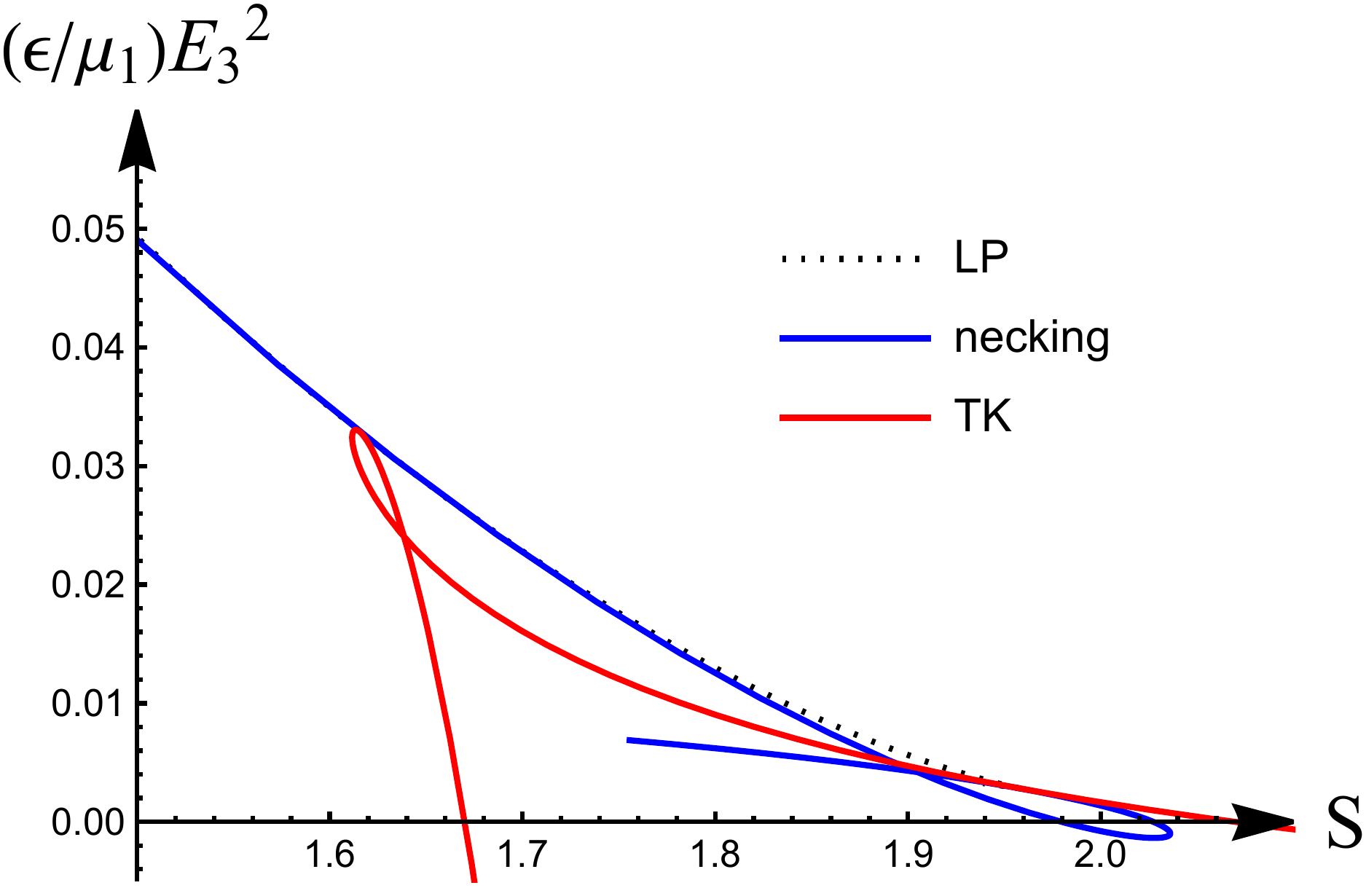}  \\ \\
(a) & & (b)
\end{tabular}
\caption{Bifurcation conditions for the TK, limiting point and necking instabilities corresponding to the strain energy function \rr{energy_2terms}. The alternative representations in (b) are obtained by viewing $E_3$ and $S$ as functions of $\lambda$ and varying $\lambda$ in the interval $(1, 3.7)$. The three lines in (a) intersect at $\lambda=1.98$ and $3.23$, and the curve associated with necking cuts the horizontal axis at $\lambda=2.44$ and $2.92$. In (b) the dotted line corresponding to the limiting point instability is close but always above that for the necking instability.}
\label{fig1}
\end{center}
\end{figure}
Corresponding to the strain energy function \rr{energy_2terms}, the three bifurcation conditions are shown in Fig.\ref{fig1}(a,b) by viewing $E_3$ as a function of $\lambda$ or $S$, respectively. Fig.\ref{fig1} (b) is obtained by eliminating $E_3$ from $S=S(\lambda, \lambda, E_3)$ using the bifurcation conditions so that both $S$ and $E_3$ are parametric functions of $\lambda$.

 In the absence of an electric field ($E_3=0$), there are two bifurcation values for the TK instability and another two bifurcation values for necking, and limiting points do not exist. This purely mechanical case has previously been discussed in Wang et al (2022). In particular, it was shown that although the first bifurcation value for the TK instability is smaller than the first bifurcation value for necking, necking can still occur first when the membrane is stretched under edge displacement control since in this case the TK instability will be suppressed.

When an electric field is applied ($E_3\ne0$), we consider two typical loading scenarios. One is to first stretch the membrane to a specified value of $\lambda$, say $\lambda=2$, in the absence of an electric field,  and then increase the electric field from zero with the edge fixed. This loading scenario corresponds to displacement control and so TK instability is suppressed. Referring to Fig.\ref{fig1} (a), this means that the first instability experienced by the membrane is the necking instability although the loading path crosses the TK instability curve.

The other loading scenario is to first increase the nominal stress $S$ to a specified value, say $1.5$, in the absence of an electric field,  and then increase the electric field from zero with $S$ fixed as a dead load. This is the loading scenario adopted by \citep{HLF2012}. Fig.\ref{fig1} (b) shows that again the first instability experienced by the membrane is the necking instability.

\section{Weakly nonlinear analysis}
\setcounter{equation}{0}
The linear analysis in the previous section only provides a necessary condition for necking. Whether a necking solution really bifurcates from the homogeneous solution or not can only be answered by a near-critical nonlinear analysis.

 To fix ideas, we may assume that the strain energy function is given by \rr{energy_2terms} and the case to be studied is when $\lambda$ is fixed in the interval $(1.98, 2.44)$ and $E_3$ is gradually increased from zero. As pointed out in the previous section, in this parameter regime necking would occur before the limiting point instability or the TK instability.

We define a non-dimensional load parameter $\omega$ through
\be \omega =\frac{\vep}{\mu_1} E_3^2. \la{omega} \en
%where $\mu$ is a typical modulus that is used to scale the strain energy function and the stresses, and is taken to be $\mu_1$ when $W$ is given by \rr{energy_2terms}.
Denoting its bifurcation value by $\omega_{\rm cr}$ (which depends on $\lambda$), we write
\be
\omega =\omega_{\rm cr}+\ep \omega_1, \la{wa1} \en
%where $\ep$ is a small positive perturbation parameter.
where $\omega_1$ is an $O(1)$ constant and $\ep$ is a positive small parameter characterizing the derivation of $\omega$ from $\omega_{\rm cr}$.
From the bifurcation condition \rr{asy5} it can be deduced that in this parameter regime the buckling mode will have $k=O(\sqrt{\ep})$, which means that the dependence of the near-critical solution on $r$ should be through the stretched variable $s$ defined by
\be s=  \sqrt{\ep} r. \la{farz} \en
The relative orders of $u, v, p^*$ and $\psi$ can be deduced by expanding the linear solutions \rr{add1} for small $k$. The absolute size of $v$ is determined by the fact that the amplitude is expected to be a linear function of $\omega -\omega_{\rm cr}$ for the type of bifurcations under consideration. This gives $v=O(\ep)$. Based on this analysis, we look for a near-critical solution of the form
$$   u=\sqrt{\ep} \left\{ u^{(1)}(s, z)+\ep u^{(2)}(s, z)+\ep^2 u^{(3)}(s, z)+\cdots \right\}, $$
\be v= {\ep} \left\{ v^{(1)}(s, z)+\ep v^{(2)}(s, z)+\ep^2 v^{(3)}(s, z)+\cdots \right\}, \la{asymsol} \en
$$   p^*=\ep \left\{ p^{(1)}(s, z)+\ep p^{(2)}(s, z)+\ep^2 p^{(3)}(s, z)+\cdots \right\}, $$
$$ \psi=\ep^2 \left\{ \psi^{(1)}(s, z)+\ep \psi^{(2)}(s, z)+\ep^2 \psi^{(3)}(s, z)+\cdots \right\}, $$
where all the functions on the right hand sides are to be determined from successive approximations.

To ease descriptions, we scale all the governing equations and boundary conditions so that the left hand side of each equation becomes of $O(1)$. This is achieved by
dividing by $\ep$ the electric equilibrium equation
\rr{aug0}$_2$, the mechanical equilibrium equation \rr{incre1}$_2$, the incompressibility condition \rr{incom2} and the boundary condition \rr{incre5a}$_1$, and by $\sqrt{\ep}$ the mechanical equilibrium equation \rr{incre1}$_1$ and boundary condition \rr{incre5a}$_2$.  On substituting \rr{asymsol} into these scaled equations and then equating the coefficients of like powers of $\ep$, we obtain a hierarchy of boundary value problems. In the following description, the two equilibrium equations in \rr{incre1} are referred to as the $r$- and $z$-equilibrium equations, respectively. At the $n$-th order ($n=1, 2$ or $3$), we integrate the $r$-equilibrium equation subject to the boundary condition \rr{incre5a}$_2$ to find  $u^{(n)}(s, z)$, the  incompressibility condition to find  $v^{(n)}(s, z)$,
and finally the $z$-equilibrium equation subject to \rr{incre5a}$_1$ to find $ p^{(n)}(s, z)$.

At leading order, the above procedure yields
\be
u^{(1)}(s, z)=A(s), \;\;\;\;
v^{(1)}(s, z)=-z \frac{1}{s} (s A(s))'+B(s), \la{4.6} \en
\be p^{(1)}(s, z)=-(\mathcal{A}^{(1)}_{3333}-\mathcal{A}^{(1)}_{2233}+\mathcal{A}^{(1)}_{3232}-\mathcal{A}^{(1)}_{3223}) \frac{1}{s} (s A(s))', \la{4.7} \en
where $A(s)$ and $B(s)$ are functions to be determined, and here and hereafter in this section all the moduli are evaluated at $\omega=\omega_{\rm cr}$. The electric equilibrium equation \rr{aug0}$_2$ is satisfied automatically.

At second order, the general solution for $u^{(2)}(s, z)$  contains two new functions $C(s)$ and $D(s)$ in the form $C(s)+z D(s)$.  Subtracting and adding the boundary condition \rr{incre5a}$_2$  at $z=\pm 1/2$, respectively, we obtain
\be
\mathcal{A}^{(1)}_{3333}-2 \mathcal{A}^{(1)}_{2233}+\mathcal{A}^{(1)}_{2222}+2 \mathcal{A}^{(1)}_{3232}-2 \mathcal{A}^{(1)}_{3223}=0, \la{4.8} \en
and
\be D(s)=- B'(s). \la{4.9} \en
The first result \rr{4.8}  is equivalent to the bifurcation condition \rr{bif}.
The general solutions for $v^{(2)}(s, z)$ and $p_2(s, z)$ contain new functions $F(s)$ and $E(s)$, respectively. On applying the boundary condition \rr{incre5a}$_1$ at $z=\pm 1/2$, we obtain $s B''(s)+B'(s)=0$, and an expression for $E(s)$. It then follows that  $B(s)=d_1 \ln s+d_2$. Since $v^{(1)}$ and hence $B(s)$ should be bounded at $s=0$,
we must set $d_1=0$. Without loss of generality we may also impose the condition $v^{(1)}(0,0)=0$ to eliminate any rigid-body displacement. This yields $d_2=0$ and hence $ B(s)=0$. Finally, integrating the electric equilibrium equation \rr{aug0}$_2$ at this order subject to \rr{incre5a}$_3$ at $z=\pm 1/2$ yields a unique expression for $\psi_1(s, z)$.
%This then completes the solution at second order.

At third order, nonlinear terms come into play and it is at this order that an amplitude equation for $A(s)$ is derived. We first solve the $r$-equilibrium equation to find an expression for $u^{(3)}(s, z)$. It contains two new functions $G(s)$ and $H(s)$ in the form $G(s)+z H(s)$. Subtracting and adding \rr{incre5a}$_2$ evaluated at $z=\pm 1/2$, respectively, we obtain the amplitude equation for $A(s)$ and an expression for $H(s)$. %It can be shown that the amplitude equation does not involve $C(s)$.
After some simplification, it is found that the amplitude equation takes the form
\be c_0 \frac{d}{ds} \frac{1}{s} \frac{d}{ds} s P'(s)+ c_1 \omega_1 P'(s)+c_2 \frac{d}{ds} P^2(s)+
c_3 A''(s) \left( A'(s)-\frac{1}{s} A(s)\right)=0, \la{4.10} \en where a prime signifies differentiation, $P(s)$ is defined by
\be P(s)=\frac{1}{s} (s A(s))', \la{4.11s} \en
and the three coefficients are given by
\begin{eqnarray}
  c_0&=&\frac{1}{12} \left(\mathcal{A}^{(1)}_{2323}-\mathcal{A}^{(1)}_{3232}\right), \nonumber \\
 c_1&=&2 \mathcal{A}^{(1)'}_{2233}+2 \mathcal{A}^{(1)'}_{2332}-\mathcal{A}^{(1)'}_{2222}-2
   \mathcal{A}^{(1)'}_{3232}-\mathcal{A}^{(1)'}_{3333}, \nonumber \\
 c_2 &=& \frac{1}{4} \left(-4 \mathcal{A}^{(1)}_{2222}-2 \mathcal{A}^{(1)}_{2233}+6 \mathcal{A}^{(1)}_{3333}-\mathcal{A}^{(2)}_{222222}+4
   \mathcal{A}^{(2)}_{222233}-\mathcal{A}^{(2)}_{112222} \right. \nonumber \\
   & & \left.-6 \mathcal{A}^{(2)}_{223333}+2 \mathcal{A}^{(2)}_{112233}+2
   \mathcal{A}^{(2)}_{333333}\right), \nonumber \\
c_3&=&\mathcal{A}^{(1)}_{2233} -\mathcal{A}^{(1)}_{2222}+\mathcal{A}^{(2)}_{222233}-\mathcal{A}^{(2)}_{112233}-\frac{1}{2}\mathcal{A}^{(2)}_{222222}+\frac{1}{2}\mathcal{A}^{(2)}_{112222}. \nonumber \end{eqnarray}
In the above expressions, $\mathcal{A}^{(1)'}_{2233}$ denotes $d \mathcal{A}^{(1)}_{2233}/d\omega$ etc., and we have used the bifurcation condition \rr{4.8} to eliminate $\mathcal{A}^{(1)}_{2332}$. It can be seen that the amplitude equation \rr{4.10} has the same structure as its mechanical counterpart derived by \cite{wjf2022}.

Corresponding to the specific free energy function \rr{energy} and \rr{energy_2terms}, we have
$$
c_0=-\frac{-480 \lambda ^{17/2}+\lambda ^{12}+800 \lambda
   ^7+3}{960 \lambda ^8}, \;\;\;\; c_1=\lambda^4, $$ \be c_2=\frac{-56 \lambda ^{17/2}+240 \lambda ^7+3}{32 \lambda ^8}, \;\;\;\;
   c_3=\frac{\sqrt{\lambda }}{2}-\frac{3 \lambda ^4}{80}. \la{ccc} \en
As a consistency check, we may neglect the nonlinear terms in \rr{4.10} to obtain
\be c_0 \frac{d}{ds} \frac{1}{s} \frac{d}{ds} s P'(s)+ c_1 \omega_1 P'(s)=0. \la{may1} \en
On substituting a solution of the form $P'(s)=J_1(k s/\sqrt{\ep})$ into \rr{may1}, where $k$ is a constant, we obtain
\be c_1 (\omega-\omega_{\rm cr}) -c_0 k^2 =0. \la{may2} \en
On the other hand, expanding \rr{asy5} around $\omega=\omega_{\rm cr}$, we obtain
\be  \left\{\frac{d}{d\omega}  (\beta +\gamma )\right\}_{\rm cr} (\omega-\omega_{\rm cr}) + \left.\frac{k^2}{24}  (\alpha-\gamma)\right|_{\rm cr}=0, \la{may3} \en
where the subscripts \lq\lq cr" signify evaluation at $\omega=\omega_{\rm cr}$. We have verified that \rr{may2} is indeed consistent with \rr{may3}.

As another consistency check, we may expand \rr{4.10} out fully and omit all the terms that are divided by powers of $s$ to obtain its {\it planar} counterpart:
\be
c_0 A^{(4)}(s)+c_1 \omega_1 A''(s)+ c_2^* A'(s) A''(s)=0, \la{2damp} \en
where
\be c_2^*=2 c_2+c_3=3 \mathcal{A}^{(1)}_{3333}-3 \mathcal{A}^{(1)}_{2222}-\mathcal{A}^{(2)}_{222222}+3 \mathcal{A}^{(2)}_{222233}-3
   \mathcal{A}^{(2)}_{223333}+\mathcal{A}^{(2)}_{333333}. \la{2dc2} \en
It has an exact solution given by
\be A(s)=\frac{6 c_0}{c_2^*} \sqrt{\frac{-c_1 \omega_1}{c_0}} {\rm tanh} \left( \frac{1}{2} \sqrt{\frac{-c_1 \omega_1}{c_0}} s\right). \la{plane} \en
This solution has the property $A'(s) \to 0$ as $s \to \infty$ and is the localised necking solution in the 2D case \citep{fdx2018}.

It does not seem possible to find a similar analytical solution for the original amplitude equation \rr{4.10} that is fourth-order with variable coefficients.
We thus resort to finding its numerical solution with the use of the finite difference method. With the use of the substitution
$ A(s) \to (c_0/c_2) \kappa^2 A(\kappa s)$, equation \rr{4.10} may be reduced to \be  \frac{d}{dt} \frac{1}{t} \frac{d}{dt} t P'(t)- P'(t)+  \frac{d}{dt} P^2(t)+
\frac{c_3}{c_2} A''(t) \left( A'(t)-\frac{1}{t} A(t)\right)=0, \la{4.10s} \en
where $t=\kappa s, \; \kappa=\sqrt{- c_1 \omega_1/c_0}$ and $P(t)$ is still defined by \rr{4.11s}.

We replace the semi-infinite interval $[0, \infty)$ by a finite interval $[0, L]$ and discretize the latter into $N$ equal intervals with node points
$$ t_i=i \tilde{h}, \;\;\;\; \tilde{h}=\frac{L}{N}, \;\;\;\; i=0, 1, 2, ..., N. $$
We apply the central finite difference scheme such that
\be
A'(t_i)=\frac{A_{i+1}-A_{i-1}}{2 \tilde{h} }, \;\;\;\; A''(t_i)=\frac{A_{i+1}-2 A_i+A_{i-1}}{\tilde{h}^2 }, \la{fd1} \en \be
 A'''(t_i)=\frac{A_{i+2}-2 A_{i+1}+2 A_{i-1}-A_{i-2}}{2 \tilde{h}^3 }, \la{fd1a} \en \be A^{(4)}(t_i)=\frac{A_{i+2}-4 A_{i+1}+6 A_i-4 A_{i-1}+A_{i-2}} {\tilde{h}^4 },\la{fd2} \en
where $A_i=A(t_i)$, etc. Evaluating the amplitude equation \rr{4.10} at the $N-1$ interior nodes $t_1, t_2, ..., t_{N-1}$, we obtain $N-1$ equations that involve the $N+3$ unknowns $A_{-1}, A_0, ..., $ and $A_{N+1}$. The remaining four equations are obtained as follows.

First, it follows from the symmetry conditions $$\lim_{s \to 0} u^{(1)}(s, z) =0 \;\;\;\; {\rm and} \;\; \lim_{s \to 0} \frac{\paa v^{(1)}}{\paa s}(s, \frac{1}{2}) =0$$ that  $\lim_{t \to 0} A(t)=0$
and $\lim_{t \to 0} P'(t)=0$. By trying a series solution for small $t$, it is found that the unique solution that satisfies the above conditions has the behavior $A(t)\sim a_1 t+a_2 t^3+\cdots$ for some constants $a_1$ and $a_2$.  This gives $\lim_{t \to 0} A''(t)=0$. The two conditions $A(0)=A''(0)=0$ together with \rr{fd1}$_2$ then yield two additional equations.

Next, we consider the asymptotic behaviour of the solutions as $t \to \infty$. Although the planar solution \rr{plane} does not decay, we expect that the solution of \rr{4.10} will experience algebraic decay due to geometric spreading. Since quadratic terms are expected to decay faster than linear terms, the decay behavior may be captured by neglecting the nonlinear terms:
\be   \frac{d}{dt} \frac{1}{t} \frac{d}{dt} t P'(t)- P'(t)=0,\;\;\;\; {\rm as}\;\; t \to \infty. \la{4.10f} \en
 The unique decaying solution of \rr{4.10f} is given by
\be P(t)=P_\infty(t) \equiv a_3 K_0( t), \;\;\;\; A(t)=A_\infty(t) \equiv \frac{a_4}{t}-a_3 K_1(  t), \la{march3} \en
where $a_3$ and $s_4$ are constants, and $K_0$ and $K_1$ are the modified Bessel function of the second kind that has the asymptotic behaviour
\be
K_\alpha(x) \sim  \sqrt{\frac{\pi}{2 x}} {\rm e}^{-x}  \left[1+\frac{4 \alpha^2-1}{8x}+\cdots \right], \;\;\;\;{\rm as}\;\; x \to \infty. \la{k0} \en
%\be \K_1(x) \sim  \sqrt{\frac{\pi}{2 x}}  \left[1+\frac{3}{8x}+\cdots \right] \,{\rm e}^{-x}, \;\;\;\;{\rm as}\;\; x \to \infty. \la{k1} \en
The asymptotic behaviour \rr{march3} is consistent with our earlier assumption that $A(s)$ decays algebraically. We note that the above decay behaviour is based on the assumption that $t$ is a real variable, or equivalently $\kappa$ is a real constant. This enables us to deduce that whenever a necking bifurcation takes place, it is generally subcritical ($\omega_1<0$ since $c_1/c_0>0$).

If a function $f(x)$ decays exponentially like ${\rm e}^{-a x}$ as $x \to \infty$ for some positive constant $a$, then it is preferable to impose the \lq\lq soft" asymptotic condition $f'(L)+a f(L)=0$ instead of the \lq\lq hard" condition $f(L)=0$ (since $f'(L)+a f(L)$ is much smaller than $f(L)$). Extending this idea, we use \rr{march3} to find the first three derivatives of $A_\infty(s)$ and by eliminating $a_1$ and $a_2$ express $A_\infty''(s)$ and $A_\infty'''(s)$ in terms of $A_\infty(s)$ and $A_\infty'(s)$.
Replacing $A_\infty(s)$ by $A(s)$ and evaluating these two expressions at $s=s_N=L$ followed by the use of \rr{fd1}--\rr{fd2}, we obtain two more additional equations. The system of $N+3$ quadratic equations can then be solved provided an appropriate initial guess is given. It is found that one good initial guess is the planar solution \rr{plane} divided by $1+s$.
\begin{figure}[ht]
\begin{center}
\includegraphics[width=.6\textwidth]{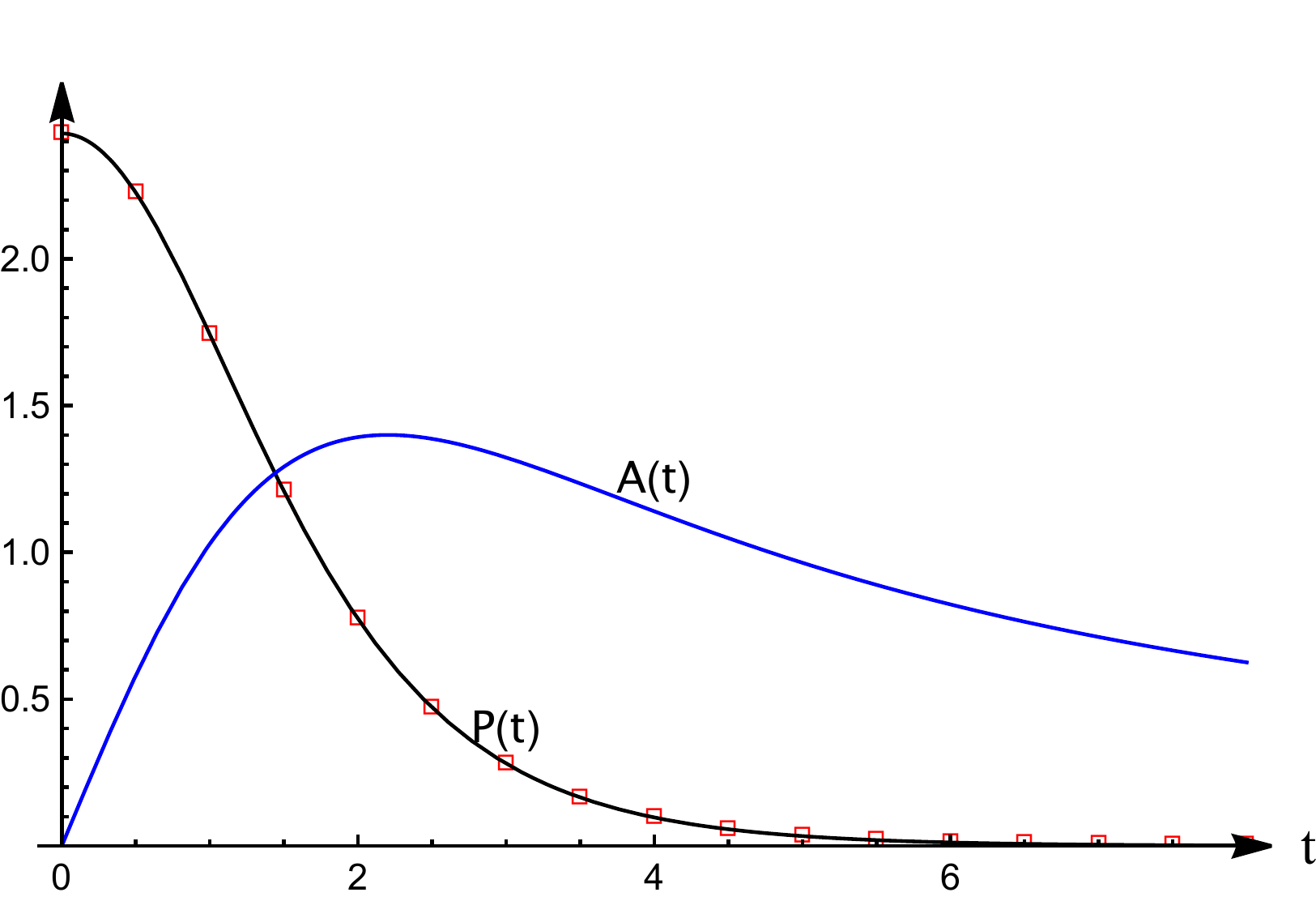}
\caption{FD solution of the amplitude equation \rr{4.10} when $\lambda=2$ and the strain energy function is given by \rr{energy_2terms}. }
\label{fig2}
\end{center}
\end{figure}
For values of $\lambda$ in the interval $(1.979, 2.439)$, it is found that the coefficient $c_3/c_2$ is positive when $\lambda<2.096$ and negative when $\lambda>2.096$. So we consider two representative cases corresponding to $\lambda=2$ and $2.2$, respectively.  It is found that taking $N=1000$ and $L=10$ yields sufficiently accurate results. Fig.\ref{fig2} shows the finite difference solution corresponding to $\lambda=2$ together with the approximate analytical solution
\be P(t)=\frac{a}{b t^2+1} {\rm sech}^2(c t), \la{appr} \en
where the constants $a, b, c$ are determined by fitting \rr{appr} to the finite difference  solution. The maximum relative error over the entire interval is less than 3.4\%. The solution corresponding to $\lambda=2.2$ for which the $c_3/c_2$ is of opposite sign is very similar and is thus not displayed here.

\section{Discussion and conclusion}
\setcounter{equation}{0}
Pull-in failure in dielectric elastomer actuators is widely believed to be associated with the limiting-point behaviour whereby the electric field as a function of the electric displacement or stretch has a maximum. For the plane-strain or plane-stress case, this connection is well explained using the analogy with the inflation problem associated with a rubber tube where the limiting-point behaviour is well-known to be associated with localised bulging that eventually evolves into a \lq\lq two-phase" state \citep{fdx2018, HS12}. However, for the case of equibiaxial tension, this explanation contradicts the fact that at large values of dead load, the limiting-point behaviour may disappear but pull-in failure can still be observed \citep{HLF2012}. Our current paper offers an alternative explanation, namely that pull-in failure evolves from axisymmetric necking through an unstable process. We note that the condition for necking does not necessarily require limiting-point behaviour.

We have only carried out a linear and weakly nonlinear analysis in the current study, but the fully nonlinear numerical simulations carried out in our earlier paper  \citep{wjf2022} for the purely mechanical case should also be indicative of what might be expected in the current electroelastic case. Thus, combining the weakly nonlinear results in the previous section with the fully nonlinear simulation results in \cite{wjf2022}, we may draw the following conclusions. When the bifurcation condition \rr{necking-condition} is satisfied, a necking solution will bifurcate from the homogeneous solution subcritically. If the membrane is gradually pulled further in the radial direction at the edge, with the electric potential fixed, the necking solution will grow in amplitude, corresponding to an increased reduction in thickness at the origin, and when a maximum amplitude is approached, the necking solution will start to propagate in the radial direction in the form of a \lq\lq two-phase" deformation. This is very similar to the localised bulging of an inflated rubber tube except that here the propagation is also accompanied by algebraic decaying of the amplitude due to geometrical spreading. On the other hand, if the electric potential is increased further from its bifurcation value while the membrane edge is fixed, the membrane will snap to a \lq\lq two-phase" deformation. This is analogous to the pressure control case in the tube inflation problem.

We wish to highlight the fact that the predictions that can be made are sensitive to the material model used. To fix ideas, we have used the strain energy function \rr{energy_2terms} as an example. To show how our results depend on the strain energy function used, we have shown in Fig.~4 the counterpart of Fig.~2 when the following Gent and Mooney-Rivlin material models are used:
\be W=-\frac{1}{2} \mu J_m \ln
(1-\frac{\lambda_1^2+\lambda_2^2+\lambda_3^2-3}{J_m}),   \la{gent} \en
\be W=\frac{1}{2} \mu \left\{ \lambda_1^2+\lambda_2^2+\lambda_3^2-3+\gamma (\lambda_1^{-2}+\lambda_2^{-2}+\lambda_3^{-2}-3) \right\}. \la{mooney} \en
\begin{figure}[ht]
\begin{center}
\begin{tabular}{ccc}
\includegraphics[scale=0.45]{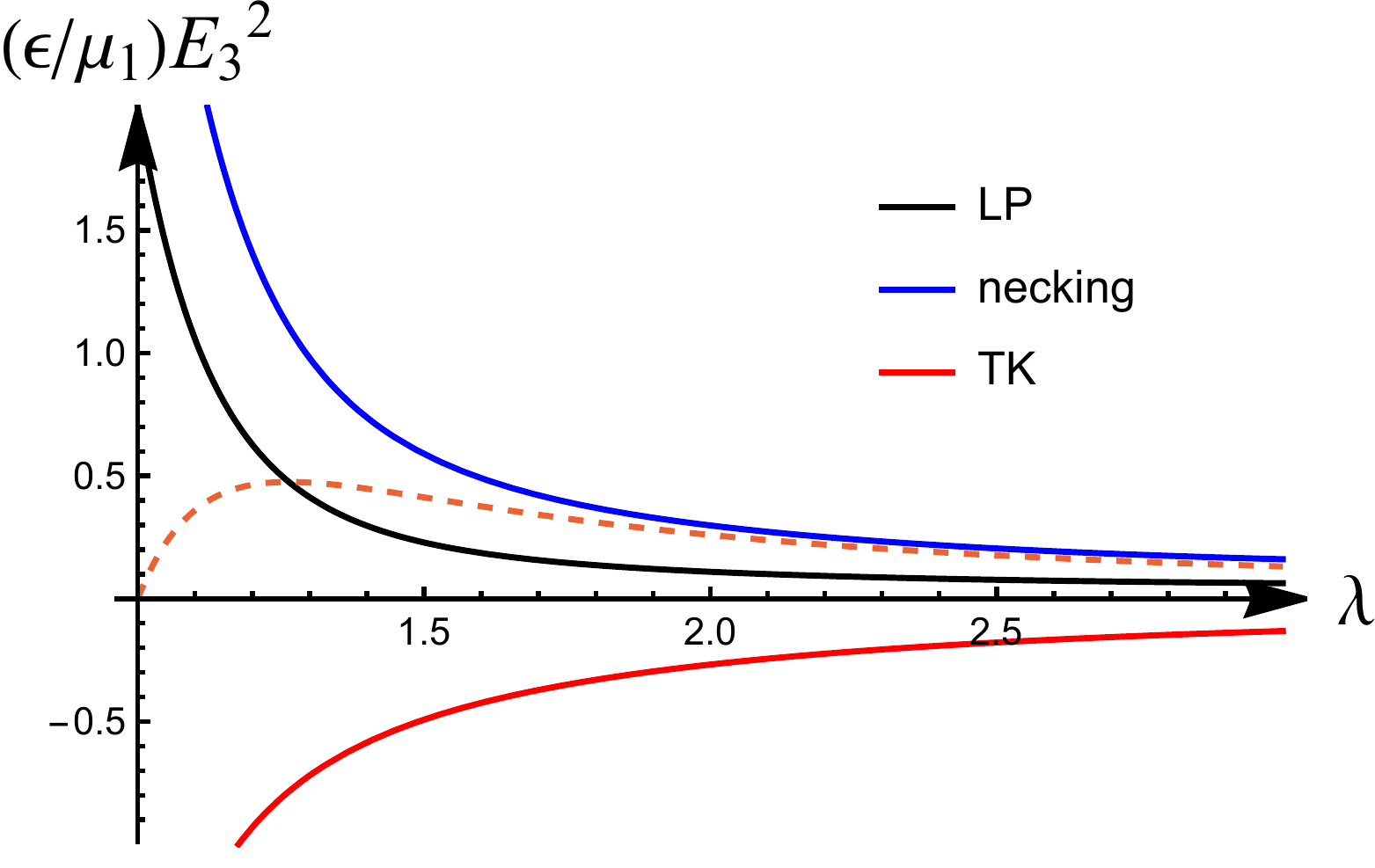}& & \includegraphics[scale=0.45]{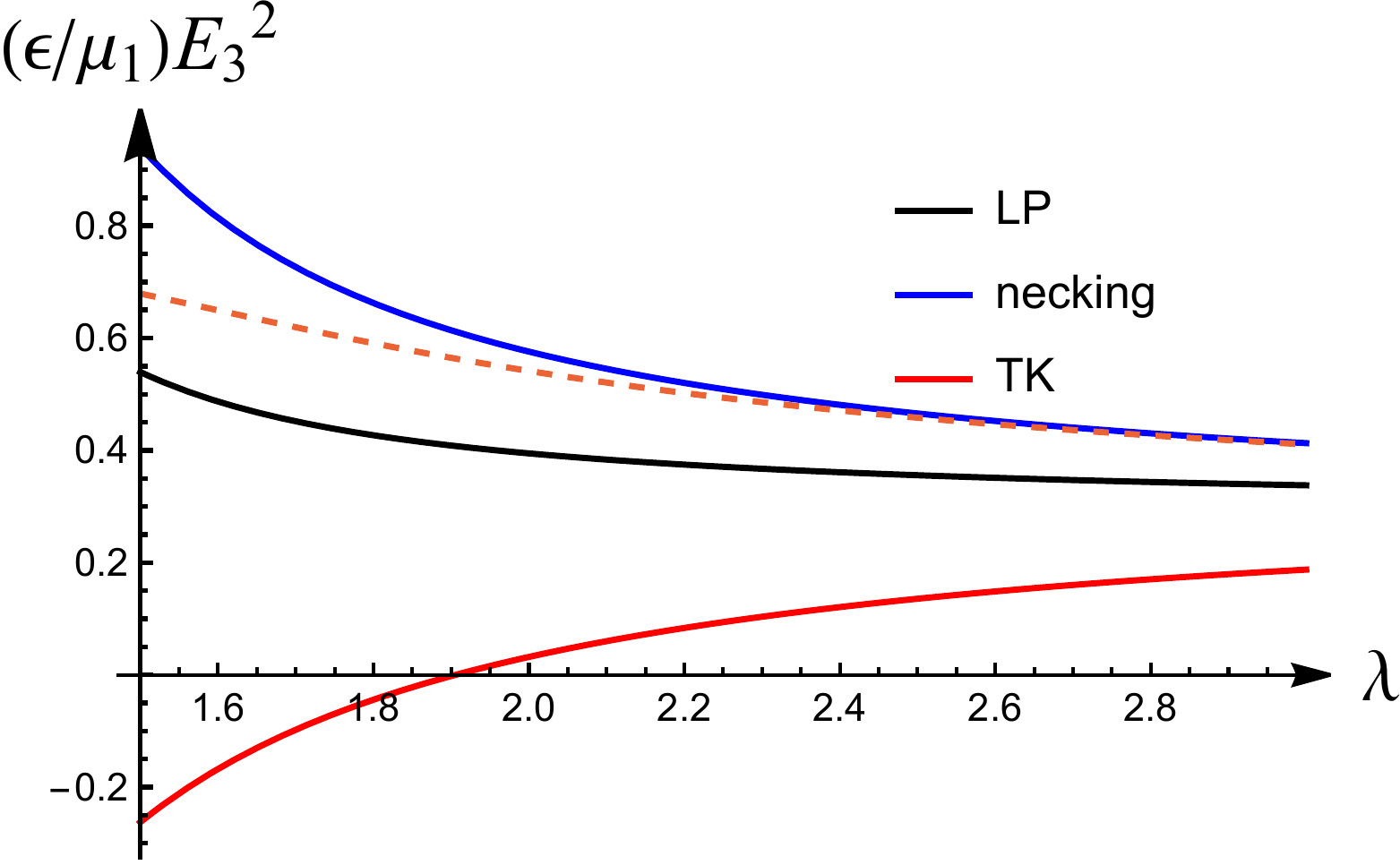} \\
(a) & & (b)
\end{tabular}
\caption{Bifurcation conditions for the TK, limiting point and necking instabilities corresponding to (a) the Gent strain energy function with $J_m=97.2$, and (b) the Mooney-Rivlin strain energy function with $\gamma=0.3$. The dashed line corresponds to zero nominal stress in the radial direction above which the nominal stress is negative.}
\label{fig3}
\end{center}
\end{figure}
It is found that the bifurcation curves have a very weak dependence on the value of $J_m$ and the curves corresponding to $J_m=\infty$ (the neo-Hookean model) are almost the same as those in Fig.~4(a) for $J_m=97.2$. It is seen that the main effect of increasing the $\gamma$ in \rr{mooney} is to shift the curves for the TK and limiting instabilities upwards. As a result, the TK instability is not possible for the Gent and neo-Hookean material models (since the corresponding $E_3$ is negative) but is possible for the Mooney-Rivlin material model. This is well-known in the purely mechanical case. The bifurcation curve for necking is always above the curve corresponding to zero nominal stress in the radial direction (dashed line). Thus,  although necking is theoretically possible, it is unlikely to be observable when the dielectric membrane has the constitutive behaviour modelled by these two material models. It then remains an open question whether there exist dielectric materials whose constitutive behaviour allows the type of axisymmetric necking that is described in the current paper. It is hoped that this question will be answered in our future experimental studies.
\section*{Acknowledgement}
%%%%%%%%%%%%%%%%%%%%%%%%%%%%%%%%%%%%%%%%%%%%%%%%%%%%%%%%%%%%%%%%%%%%%%%%%%
This work was supported by the National Natural Science Foundation of China (Grant No 12072224) and the Engineering and Physical Sciences Research Council, UK (Grant No EP/W007150/1).

\end{document}